\documentclass[]{aa}  

\usepackage{natbib}
\usepackage{multirow}
\bibpunct{(}{)}{;}{a}{}{,}
\usepackage{graphicx}
\usepackage{xcolor}
\usepackage[varg]{txfonts}
\begin{document} 

   \title{Oxygen abundances in G- and F-type stars from HARPS}
   \subtitle{Comparison of [OI] 6300\AA~ and OI 6158\AA~}

   \author{S. Bertran de Lis\inst{1,2}, E. Delgado Mena\inst{3}, V. Zh. Adibekyan\inst{3}, N. C. Santos\inst{3,4} \and S. G. Sousa\inst{1,3,4}
          }

   \institute{Instituto de Astrof\'isica de Canarias, E-38205 La Laguna, Tenerife, Spain\\
              \email{sbertran@iac.es}
         \and
             Universidad de La Laguna, Dept. Astrof\'isica, E-38206 La Laguna, Tenerife, Spain
         \and
         	Centro de Astrof\'isica da Universidade do Porto, Rua das Estrelas, 4150-762 Porto, Portugal
         \and
             Departamento de F\'isica e Astronomia, Faculdade de Ci\^encias da Universidade do Porto, Portugal
         }

   \date{Received ...; Accepted ...}

\titlerunning{Oxygen abundances}
\authorrunning{S. Bertran de Lis et al.}
% \textcolor{red}{}
% \abstract{}{}{}{}{} 
% 5 {} token are mandatory
  \abstract
  % context heading (optional) 
%{The reliability of oxygen abundance measurements from the [OI]6300 line in solar-type stars has been discussed by many authors.}
{}
  % aims heading (mandatory)
{We present a detailed and uniform study of oxygen abundance from two different oxygen lines at 6158$\AA$ and 6300$\AA$ in a large sample of solar-type stars. The results are used to check the behaviour of these spectral lines as oxygen abundance indicators and to study the evolution of oxygen in thick and thin disk populations of the Galaxy.}
  % methods heading (mandatory)
{Equivalent width measurements were carried out for the [OI]~6158$\AA$ and OI~6300$\AA$ lines. LTE abundances were obtained from these two lines in 610 and 535 stars, respectively. We were able to measure oxygen abundance from both indicators in 447 stars, enabling us, for the first time, to compare them in a uniform way. Careful error analysis has been performed.}
     % results heading (mandatory)
{We found that oxygen abundances derived from the 6158$\AA$ and 6300$\AA$ lines agree to within 0.1dex in 58\% of the stars in our sample, and this result improves for higher signal-to-noise values. We confirm an oxygen enhancement in stars of the thick disk, as has also been seen for other $\alpha$-elements. The new oxygen abundances confirm previous findings for a progressive linear rise in the oxygen-to-iron ratio with a slope equal to 0.78 from solar metallicity to [Fe/H]$\sim$-1. However, the slope we measured is steeper than the one found in previous studies based on the oxygen triplet. Below [Fe/H]=$-$0.6 our stars show [O/Fe] ratios as high as $\sim$0.8, which can be interpreted as evidence for oxygen overproduction in the Galactic thick disk. These high oxygen abundances do not pose a problem to chemodynamical models since there is a range of parameters that can accommodate our results.}
	% conclusions (optional)
{}
  

   \maketitle

%%%%%%%%%%%%%%%%%%%%%%%%%%%%%%%%%%%%%%%%%%%%%%%%%%%
% INTRODUCTION
%%%%%%%%%%%%%%%%%%%%%%%%%%%%%%%%%%%%%%%%%%%%%%%%%%%

\section{Introduction}
After hydrogen and helium, oxygen is the most abundant element in the Universe. 
As a matter of fact, oxygen is the only chemical element that has a unique production site. 
This has been used by many authors to argue that oxygen is the best tracer of galactic chemical
evolution. It is produced by alpha particles during hydrostatic nucleosynthesis in massive stars. The interstellar medium is enriched with oxygen when massive stars explode as Type II supernovae.  \citep[e.g.][]{Wheeler89, Maeder09, Stasinska12}.

% Two column figure 
\begin{figure*}[ht]
   \centering
	\begin{minipage}[b]{0.33\linewidth}
		\resizebox{\columnwidth}{!}{\includegraphics{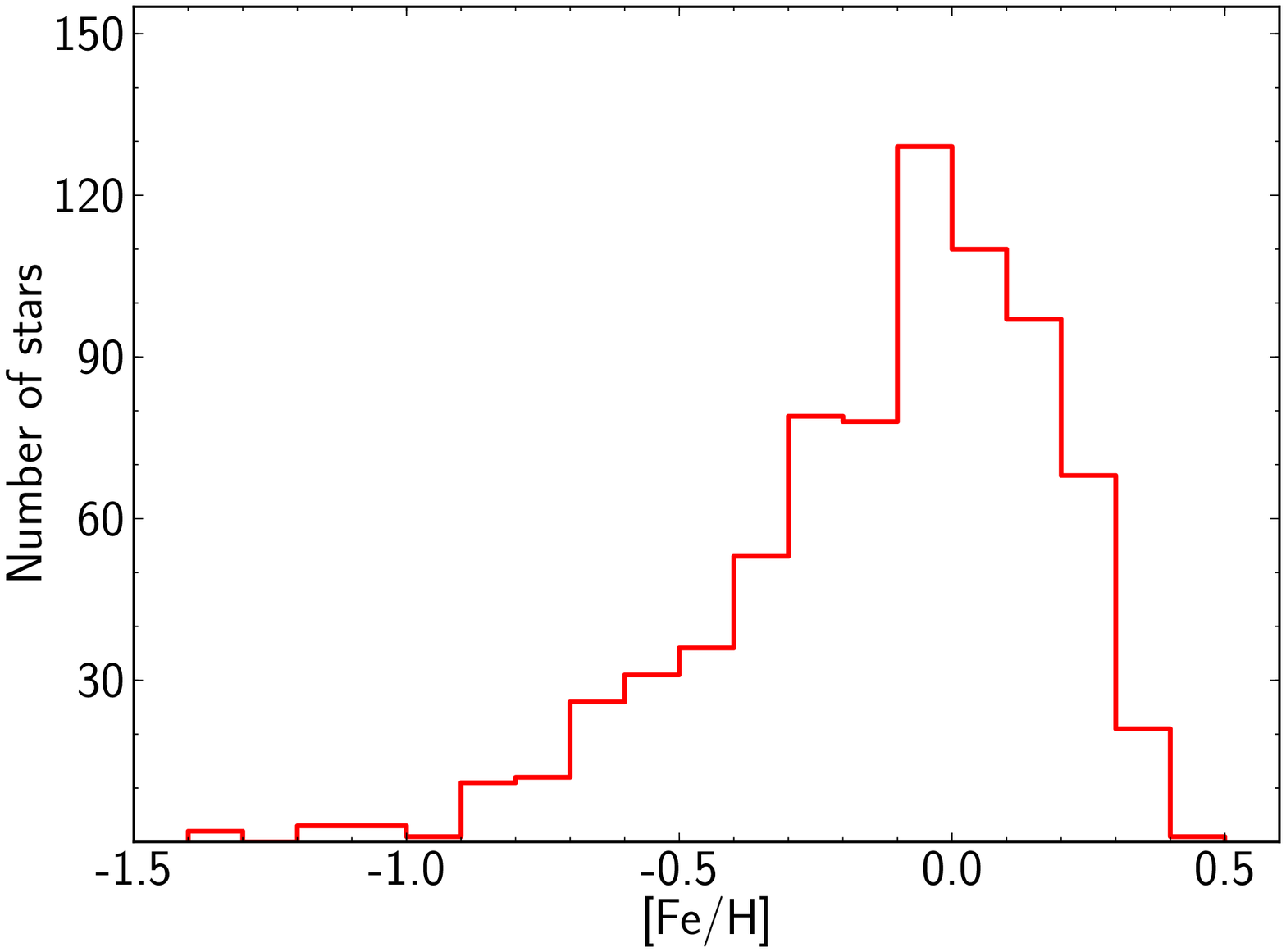}}
	\end{minipage} 
	\begin{minipage}[b]{0.33\linewidth}
		\resizebox{\columnwidth}{!}{\includegraphics{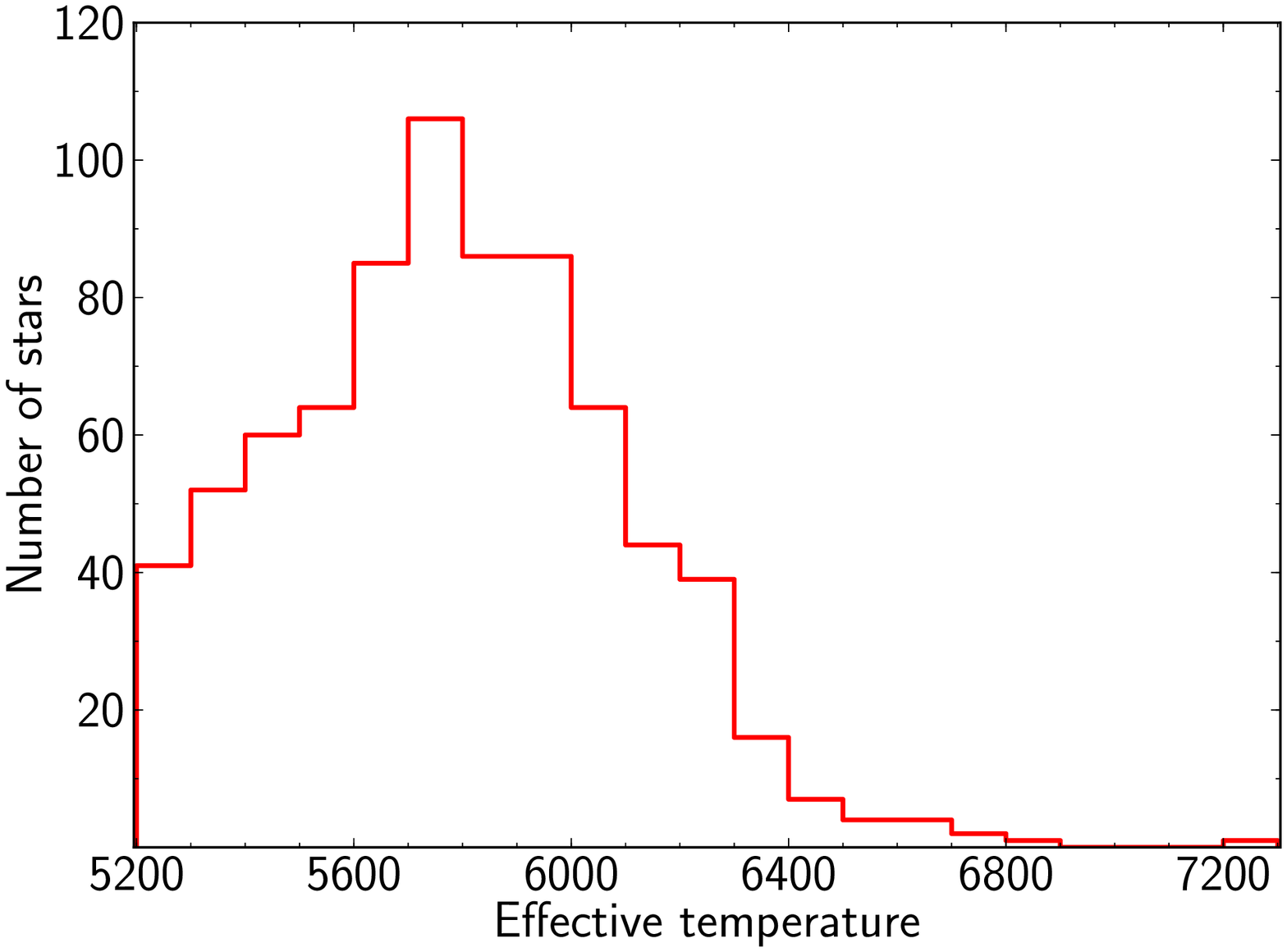}}
	\end{minipage} 
	\begin{minipage}[b]{0.33\linewidth}
		\resizebox{\columnwidth}{!}{\includegraphics{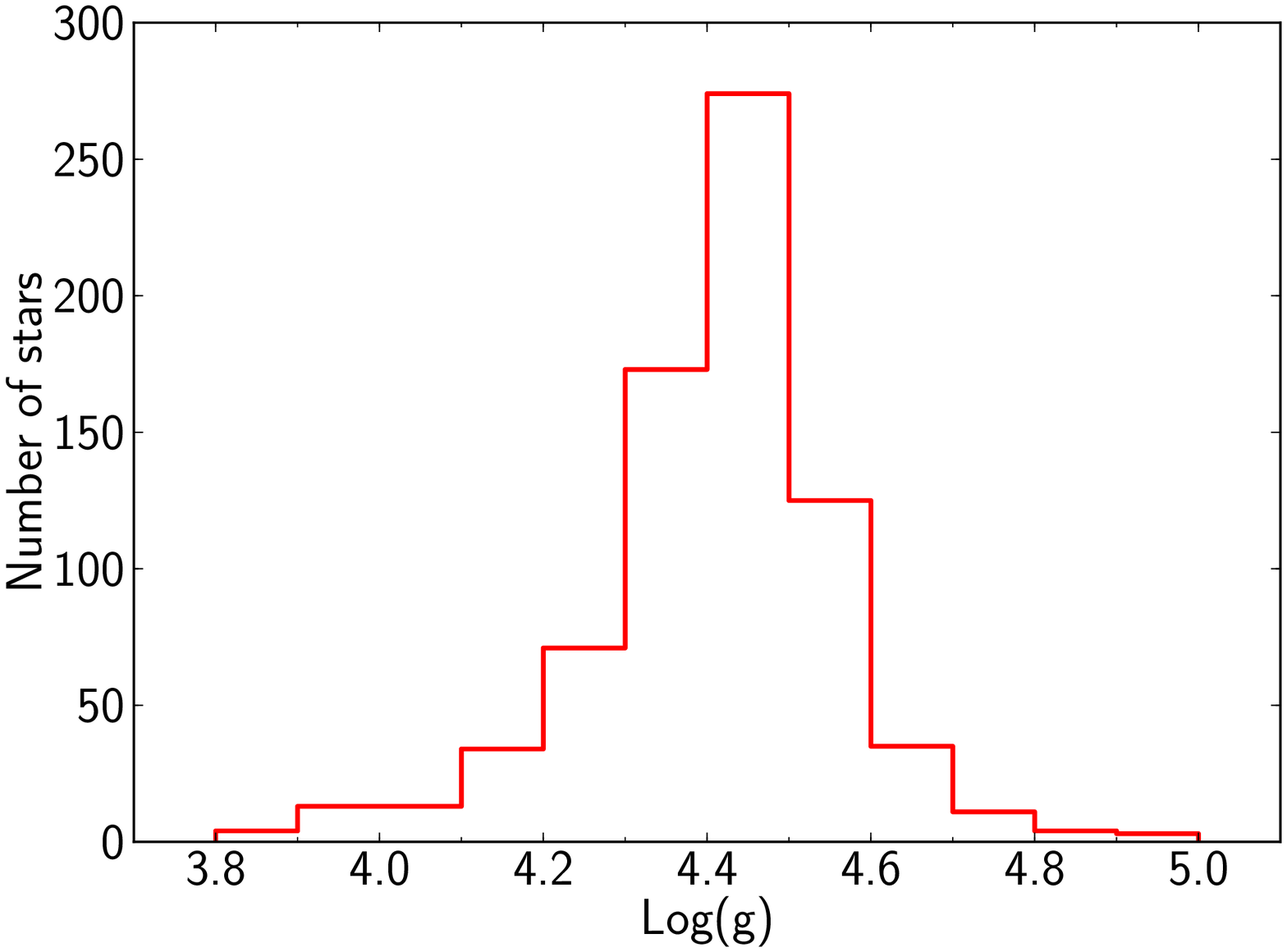}}
	\end{minipage} 
\caption{Distribution of the fundamental stellar parameters in the HARPS sample for stars with effective temperatures higher than 5200 K.}
\label{params}
\end{figure*}

Despite the importance of deriving precise oxygen abundances in solar-type stars, this is usually not a trivial process. The number of spectral lines of atomic oxygen suitable for abundance analysis in visual spectra is small. There is a large number of molecular lines of OH in the near-UV and near-IR, strong enough to be measured in main-sequence metal-poor stars down to [Fe/H]$\sim$-3.5~dex \citep{Israelian98, Israelian01, Boesgaard99}. However, it is extremely difficult to measure abundances from molecular features as they are affected by several physical processes that are not well understood in solar-type stars (UV opacities, etc.). The situation with atomic lines is no more promising. Serious disagreements exist in both solar-type main sequence and evolved sub-giant/giant stars in any metallicity range. The well-known spectral lines of atomic oxygen suitable for abundance analysis are [OI]~6300$\AA$ and the OI triplet at 7771--4$\AA$. The weak spectral lines of OI~6158$\AA$ and [OI]~6363 $\AA$ are hardly used owing to their small equivalent widths. The strong lines of the oxygen triplet are strongly affected by deviations from local thermodynamic equilibrium (LTE) \citep[e.g.][]{Caffau08}, which depend on the cross section of collisions between the atoms of oxygen and neutral hydrogen. The physical description of these collisions is still a matter of scientific debate. Moreover, the forbidden line is often blended with telluric features and also contains a Ni I blend at 6300.40$\AA$ \citep[e.g.][]{Lambert78, Allende01} and a much weaker CN blend \citep[e.g.][]{Teske13}. The Ni line, with an uncertain oscillator strength, may contribute up to 60\% to the equivalent width (EW) of the 6300$\AA$ feature (see section~\ref{SectionComp}). In addition, a disagreement between the OI triplet and forbidden line abundances in metal-poor giants can reach up to 1~dex \citep{Israelian04}.

There are many articles dedicated to oxygen measurements in solar-type stars. Several workshops and conferences have been held to review the status of oxygen measurements and address unresolved problems related to inconsistent abundances derived from different oxygen lines. Most of the previous work on oxygen with large samples of stars \citep[e.g.][]{Bensby04, Takeda05, Ecuvillon06, Delgado10, Petigura11, Ramirez13} were based on two lines: OI triplet $\sim$7774 $\AA$ and [OI]~6300$\AA$. \citet{Ecuvillon06} have compared near-UV OH lines with the OI triplet and [OI] and found good agreement between the [O/H] ratios from forbidden and OH lines, while the NLTE (non-local thermodynamic equilibrium) triplet showed a systematically lower abundance. \citet{Takeda05} have used the 6158$\AA$ OI line to derive oxygen abundances in 160 FGK dwarfs and sub-giants. Nevertheless, the quality of their data was not high enough to reach conclusions about the reliability of this line.

To all the above-mentioned we must add that the solar oxygen abundance itself is still uncertain. Since the first determinations of \citet{Lambert78} (log~$\epsilon$(O)=8.92) and \citet{Anders89} (log~$\epsilon$(O)=8.93), the value of solar oxygen has undergone several revisions that yielded progressively lower values. In recent years \citet{Asplund04} have recommended a value of log~$\epsilon$(O)=8.66$\pm$0.05 based on measurements of [OI], OI, and IR OH lines, which was supported by the \citet{Socas-Navarro07} determination (log~$\epsilon$(O)=8.63) using the infrared triplet. This value was reviewed in 2009 to log~$\epsilon$(O)=8.69$\pm$0.05 using new 3D hydrodynamical models of the solar atmosphere \citep{Asplund09}. On the other hand \citet{Ayres08} determined a significantly larger value (log~$\epsilon$(O)=8.81) based on a single snapshot of a 3D model for the [OI]6300 line and treating the log(\textit{gf}) of Ni blended line as a free parameter. Recently, \citet{Caffau08} presented a new determination of the solar photospheric oxygen abundance by analysing different spectral atlases of the solar flux and disk--centre intensity, making use of the latest generation of CO5BOLD 3D solar model atmospheres. They have studied the photospheric oxygen abundance by considering only lines from atomic transitions. Ignoring the role of collisions with hydrogen atoms on the NLTE level populations of oxygen, they proposed log~$\epsilon$(O)=8.76$\pm$0.07. They have stressed that the measurement of equivalent widths with high precision is still an important and open issue. Strangely enough, even for the solar high-resolution atlas \citep{Kurucz84}, they concluded that the placement of the continuum is not trivial, and that the influence of many blends gives rise to significant uncertainties.

Our work presents a complete and uniform study of the oxygen abundances in a large sample of stars from the HARPS survey, using two different abundance indicators: the high excitation line at 6158$\AA$ and the oxygen forbidden line at 6300$\AA$ . This is the first systematic comparison of these lines in a large sample  of main-sequence solar-type stars. The paper is organized as follows: in Sect. 2, we introduce the sample used in this work. The
method of the chemical abundance determination, analysis and uncertainties calculations will be explained in Sect. 3. Sect. 4 is dedicated to the comparison between 6158$\AA$ and 6300$\AA$ as oxygen abundance indicators, where we provide details about their reliability in different ranges of stellar parameters. A discussion of the [O/Fe] trends, separation of the different populations of stars and comparison with previous works can be found in Sect. 5. Finally, models of oxygen production in the Galaxy and comparison with observations is presented in section 6.

%%%%%%%%%%%%%%%%%%%%%%%%%%%%%%%%%%%%%%%%%%%%%%%%%%%
% SAMPLE
%%%%%%%%%%%%%%%%%%%%%%%%%%%%%%%%%%%%%%%%%%%%%%%%%%%
\section{Sample description and stellar parameters}

In the past decade a huge effort has been made to discover extrasolar planets. Several radial velocity surveys have carried out a thorough follow-up of solar-type stars over several years, resulting in homogeneous homogeneous sets of high resolution spectra and large catalogues of uniform stellar parameters \citep[e.g. SWEET-Cat --][]{Santos13}. Besides the well-known success of such projects for discovering extrasolar planets, these observations give us, for the first time, the opportunity to study the atmospheres and the chemical compositions of solar-type stars with an unprecedented high-quality, homogeneous and extended dataset.

The sample of stars used in this work is a combination of three subsamples from the HARPS planet search programme: HARPS-1 \citep{Mayor03}, HARPS-2 \citep{LoCurto10} and HARPS-4 \citep{Santos11}. These stars are mostly slowly rotating chromospherically inactive, unevolved solar-type dwarfs and some sub-giants, with spectral types between F2 and M0. The individual spectra were reduced with the HARPS pipeline and then combined using IRAF\footnote{\label{note1}IRAF is distributed by National Optical Astronomy Observatories, operated by the Association of Universities for Research in Astronomy, Inc., under contract with the National Science Foundation, USA.} after correcting for its radial velocity. The combined spectra for each star have a resolving power R $\sim$ 115000 and a signal-to-noise (S/N) between $\sim$40 and $\sim$2000. About 90\% of the stars have an S/N higher than 100. This high-quality data allowed us to measure the 6158 $\AA$ line in a large sample of stars and investigate its reliability as an oxygen abundance indicator.

We selected 762 stars from the HARPS samples with temperatures higher than 5200K, for which our stellar parameters are most precise \citep{Tsantaki13}. Stellar parameters were determined by \citet{Sousa08, Sousa11a, Sousa11b} using the same spectra that we studied. The typical precision uncertainties are of about 30 K for T$_{eff}$, 0.06~dex for log(\textit{g}), 0.08 km s$^{-1}$ for $\xi_{t}$, and 0.03~dex for [Fe/H]. Figure~\ref{params} shows the distribution of stellar parameters for our final sample.

%%%%%%%%%%%%%%%%%%%%%%%%%%%%%%%%%%%%%%%%%%%%%%%%%%%
% ABUNDANCE ANALYSIS
%%%%%%%%%%%%%%%%%%%%%%%%%%%%%%%%%%%%%%%%%%%%%%%%%%%
\section{Abundance analysis}
\label{Analysis}

\begin{figure*}[ht]
\centering
\resizebox{\hsize}{!}{\includegraphics[trim=0 70 0 130]{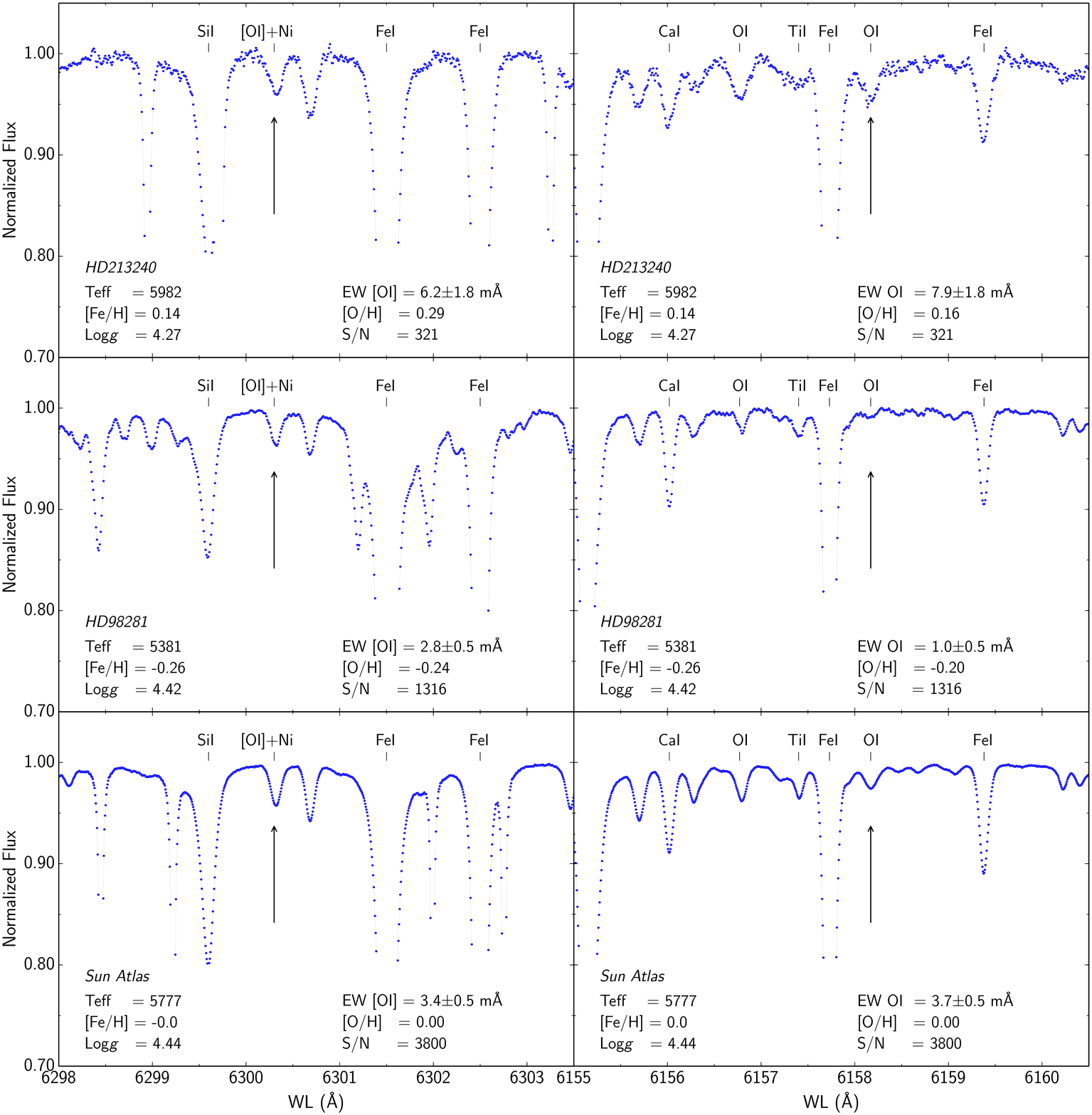}}
\caption{Spectral regions containing oxygen lines in HD213240, a star from our database with an average S/N = 300 representative for our catalogue, HD98281, for which the high S/N allows us to measure lines as small as 1 m$\AA$, and for the Kurucz Sun Atlas \citep{Kurucz84}. Stellar parameters, equivalent width of the oxygen lines (Ni blend contribution removed) and LTE abundances are provided in the plot. We note that the provided signal-to-noise ratio is the nominal S/N.}
\label{Spectra}
\end{figure*}

\begin{figure}[]
	\resizebox{\hsize}{!}{\includegraphics{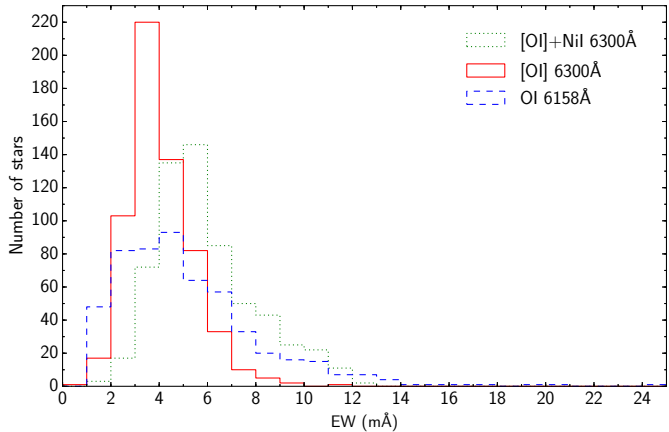}}
\caption{Distribution of measured equivalent widths for each spectral line.}
\label{EW}
\end{figure}

Oxygen abundances were derived from OI6158$\AA$ and [OI]6300$\AA$ spectral lines. There are two main issues that make the automatic EW measurement of these lines problematic. First, the two lines are placed in complex spectral regions close to some strong FeI lines and, in the case of the 6300$\AA$ transition, also with telluric lines in its vicinity (Fig.~\ref{Spectra}). Moreover, the strength of these lines in F- and G-type stars is generally smaller than 8 m$\AA$ (Fig.~\ref{EW}). Altogether, this makes the continuum placement and EW measurement a serious challenge for any automatic code. Therefore, EW measurements were carried out for both lines in a detailed individual analysis using the \textit{splot} package from IRAF\footnotemark[\value{footnote}]. 

\begin{figure*}[ht]
   \centering
	\begin{minipage}[b]{0.49\linewidth}
		\resizebox{\hsize}{!}{\includegraphics{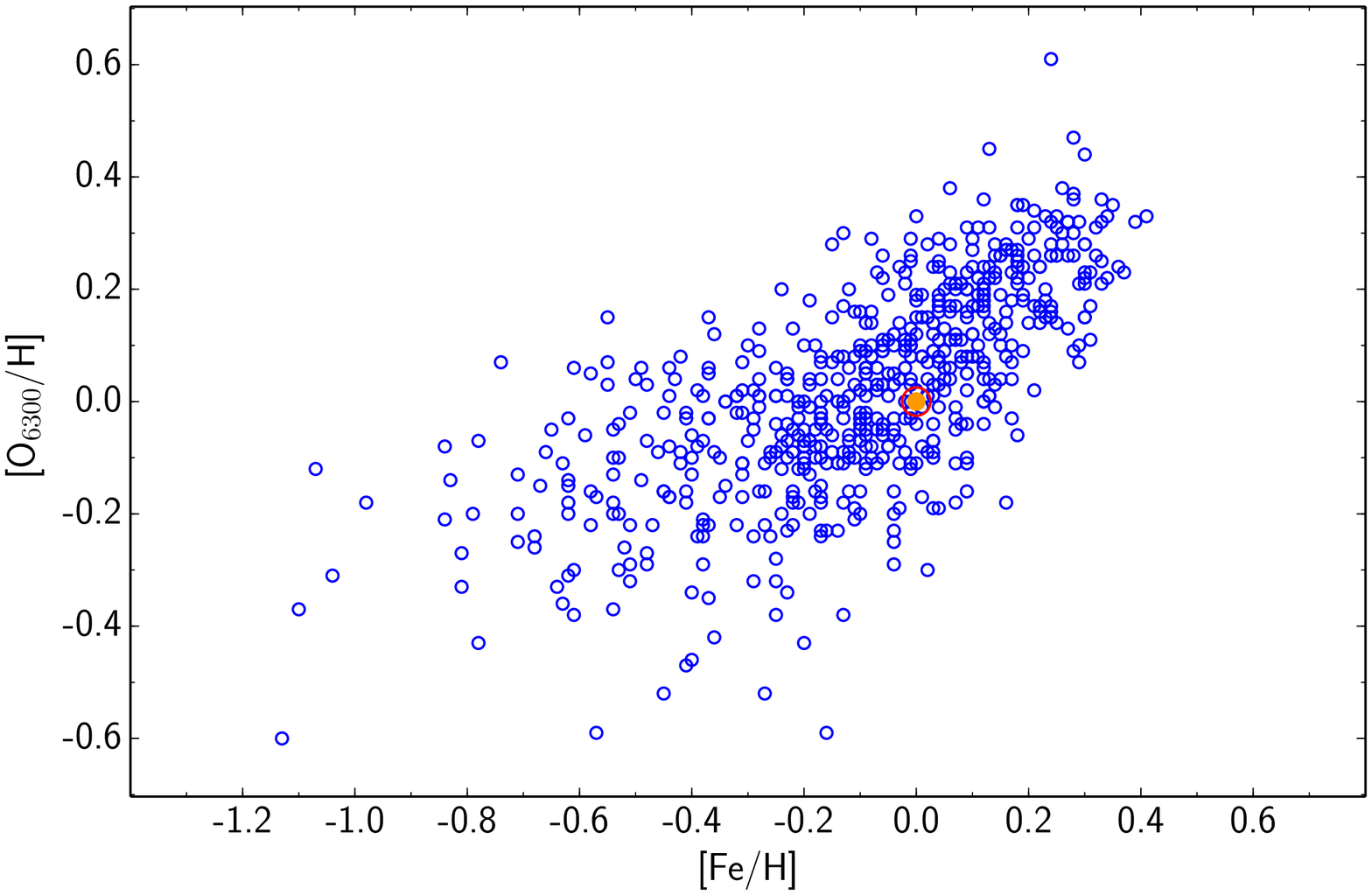}}
	\end{minipage} 
	\begin{minipage}[b]{0.49\linewidth}
		\resizebox{\hsize}{!}{\includegraphics{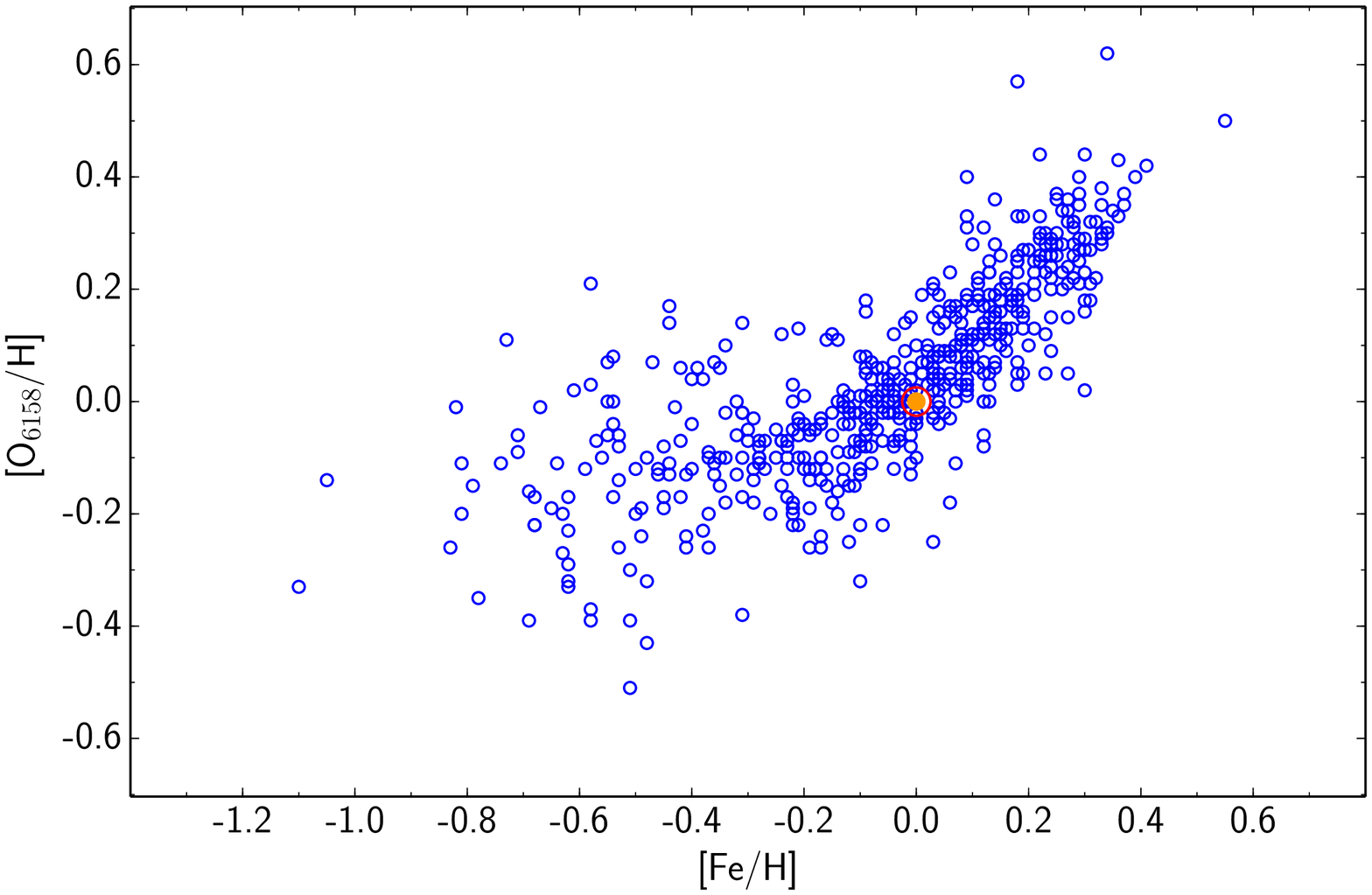}}
	\end{minipage} 
\caption{[O/H] vs.\ [Fe/H] for different indicators. The yellow dot corresponds to the solar abundances.}
\label{OH}
\end{figure*}

In combined spectra the same atmospheric lines can appear at different wavelengths, slightly shifted between them and with different intensities. To account for possible blends with telluric lines, we made a thorough observation of the spectra in the region of the forbidden line. This process was done by identifying the atmospheric pattern, which is repeated along the region of [OI]6300$\AA$, in a nearby window free from stellar lines. This pattern can be later extrapolated to check if it is superimposed to the 6300$\AA$ line. On the other hand, de-blending was required for the 6158$\AA$ feature in order to account for the presence of the FeI line at 6157.5$\AA$. The red wing of this Fe line frequently overlaps with the oxygen line. The spectral windows containing both oxygen spectral lines are shown in Figure~\ref{Spectra}. For some still unknown reasons, the nearby line of oxygen at 6156.75$\AA$ provides very different abundances from those based on the 6158$\AA$ and 6300$\AA$ lines. In fact, being even stronger than the line at 6158$\AA$, it has been avoided in previous studies \citep[e.g.][]{Caffau08, Takeda05}.

The Solar Atlas of Kurucz \citep{Kurucz84} has been used to select continuum regions free from spectral lines. We found that the continuum windows are 6156.94--6157.10$\AA$ and 6158.80--6158.90$\AA$ for OI~6158$\AA$, and 6298.20--6298.80$\AA$,  6297.15--6297.45$\AA$ and 6303.00--6303.30$\AA$ for the [OI]~6300$\AA$ line. We have also selected two stars in our sample, with temperatures far from that of the Sun, in order to verify the reliability of those regions at different T$_{eff}$. This procedure guarantees a uniform continuum placement in our targets, whose spectra are similar to those of the reference stars.

\begin{table}[]
 \caption[]{\label{AtomParams}Atomic parameters of the spectral lines together with our measurements of equivalent width and abundance in the Sun. This solar value will be used as reference in the present analysis.}
 \centering 
\begin{tabular*}{0.49\textwidth}{ccccccc}
 \hline \hline \\[-8pt]
&$\lambda$&log($gf$) &$\chi_{lo}$&Ref. &EW$_{\odot}$&log$\epsilon$(X)$_{\odot}$\\
&($\AA$)& &(eV)& &(m$\AA$)&\\[2pt]
\hline \\[-8pt]
OI& 6158.171 & -0.296 & 10.74 & 1 & 3.7 & 8.71\\
$[$OI$]$& 6300.304 & -9.717 & 0.00 & 1 & 3.4 & 8.65\\
NiI& 6300.336 & -2.110 & 4.27 & 2 & 2.0 & 6.25\\[2pt]
\hline
\end{tabular*}
\tablebib{(1)~\citet{Caffau08}, (2)~\citet{Johansson03}}
\end{table}

Oxygen abundances from two indicators were determined according to a standard LTE analysis with the revised version of the MOOG2013 spectral synthesis code \citep{Sneden73}, using the \textit{abfind} driver and a grid of Kurucz ATLAS9 plane--parallel model atmospheres \citep{Kurucz93}. Parameters of spectral lines considered in our analysis are listed in Table~\ref{AtomParams}. NLTE corrections for OI6158$\AA$ are negligible \citep[e.g.][]{Caffau08} and were not taken into account. On the other hand [OI]6300$\AA$ is not affected by deviations from LTE. The two oxygen lines have very different excitation energies, but very similar strengths in the Solar Atlas \citep{Kurucz84}. Once we move towards hotter/cooler stars along the main sequence, one of these lines will become stronger and the other weaker. Thanks to the wide range of T$_{eff}$ and [Fe/H] of the stars in our sample (Fig.~\ref{params}), we have an interesting opportunity to compare oxygen abundances derived from two lines with very different formation physics. We note that, for still unknown reasons, these lines provide different oxygen abundances in the Sun \citep{Caffau08}. One of the important questions that will be addressed in this article is whether or not this difference is seen in other sun-like stars. In general, we want to understand which of these lines is more reliable as an oxygen abundance indicator and why.

It is well known that the forbidden line at 6300.30 $\AA$ is severely blended with the NiI 6300.34$\AA$ line. The contribution of the NiI line is negligible at [Fe/H] $<$ $-$0.8, but becomes very important at solar metallicities (see section \ref{GalTrends}). The EW of the NiI line in our stars was estimated using the \textit{ewfind} driver of MOOG and Ni abundances from \citet{Adibekyan12}, calculated from the same spectra and stellar parameters used in this work. The derived EW of the Ni line was subtracted from the total EW of the 6300$\AA$ feature to obtain the contribution of oxygen. 

Many authors use the so-called solar gf-values to carry out differential abundance analysis, meaning that they force the oscillators strengths of different spectral lines of the same chemical element to provide a unique  abundance value. Here we take a different approach. We do not modify the gf-values of the oxygen lines to obtain the same abundance. In this approach we obtain 0.06~dex difference between the solar oxygen derived from the 6158$\AA$ and 6300$\AA$ lines. We want to investigate whether this difference exists in other sun-like stars with a quality of spectra similar to the Solar Atlas. The solar abundances listed in Table~\ref{AtomParams} are used as a reference in the analysis.

Abundances from OI6158$\AA$ were obtained for 535 stars, while 610 stars were analysed using [OI]6300$\AA$ from the initial sample of 762 stars. The final results from both lines are shown in Figure~\ref{OH}. These are presented relative to the solar reference abundance (Table~\ref{AtomParams}), derived following the same procedure as for the rest of the sample. Tables~\ref{O6158Results} and~\ref{O6300Results} provide the abundances from the OI6158$\AA$ and [OI]6300$\AA$ lines, respectively, together with the stellar parameters, the equivalent width of the lines and its total uncertainty. 

\begin{figure*}[ht]
   \centering
	\begin{minipage}[b]{0.49\linewidth}
		\centering
		\resizebox{\hsize}{!}{\includegraphics{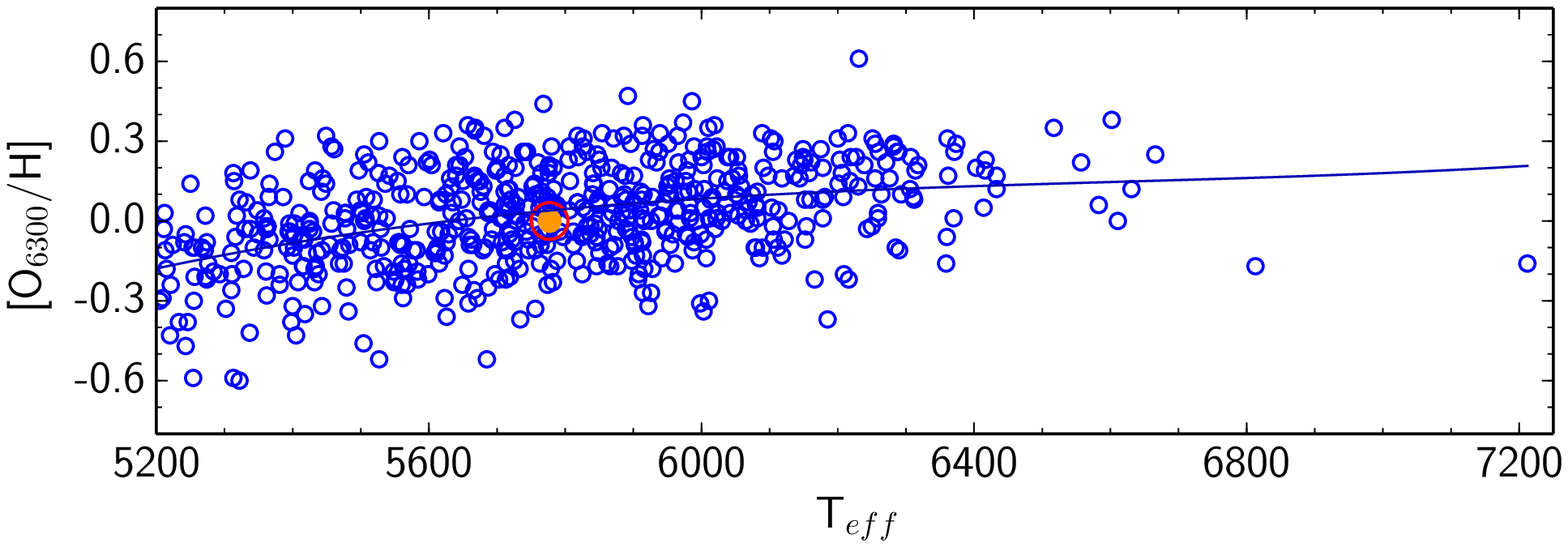}}
		\resizebox{\hsize}{!}{\includegraphics{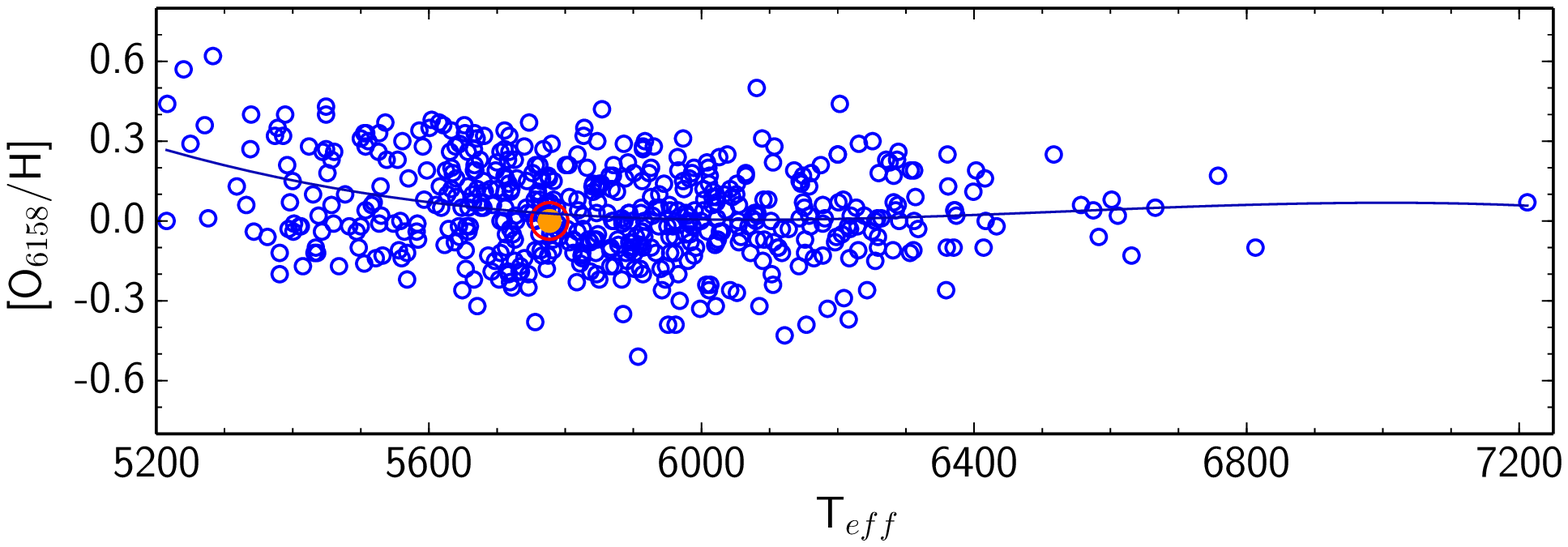}}
	\end{minipage} 
	\begin{minipage}[b]{0.49\linewidth}
		\centering
		\resizebox{\hsize}{!}{\includegraphics{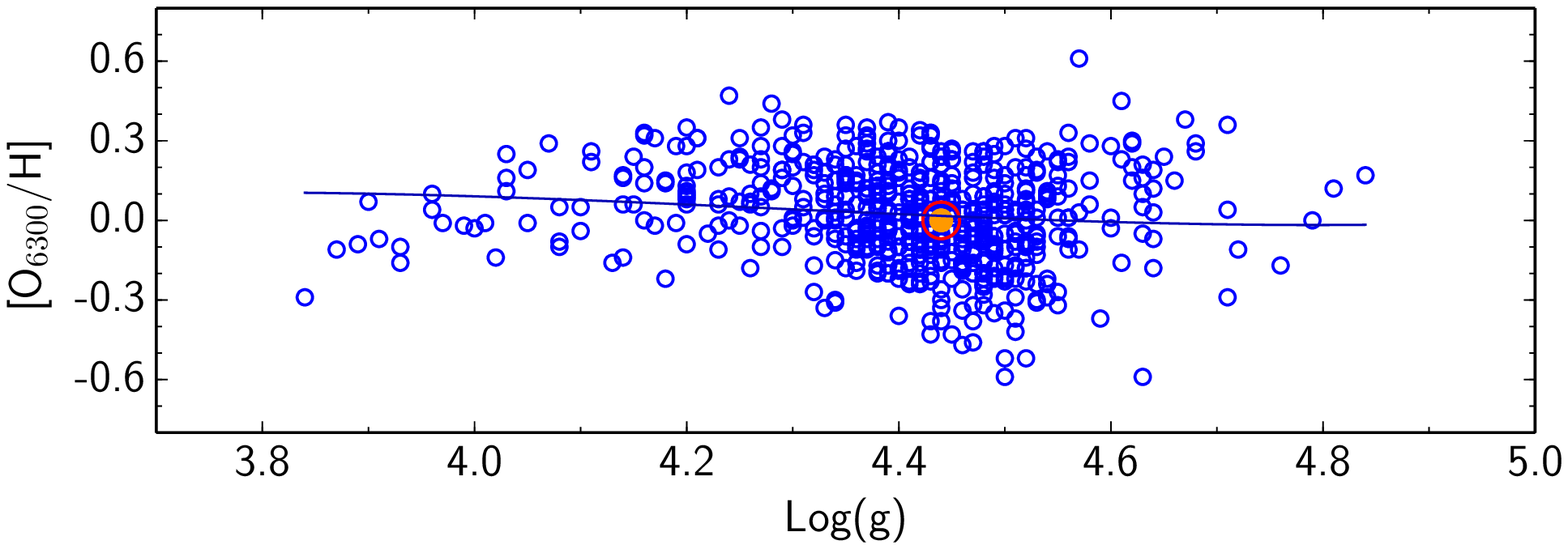}}
		\resizebox{\hsize}{!}{\includegraphics{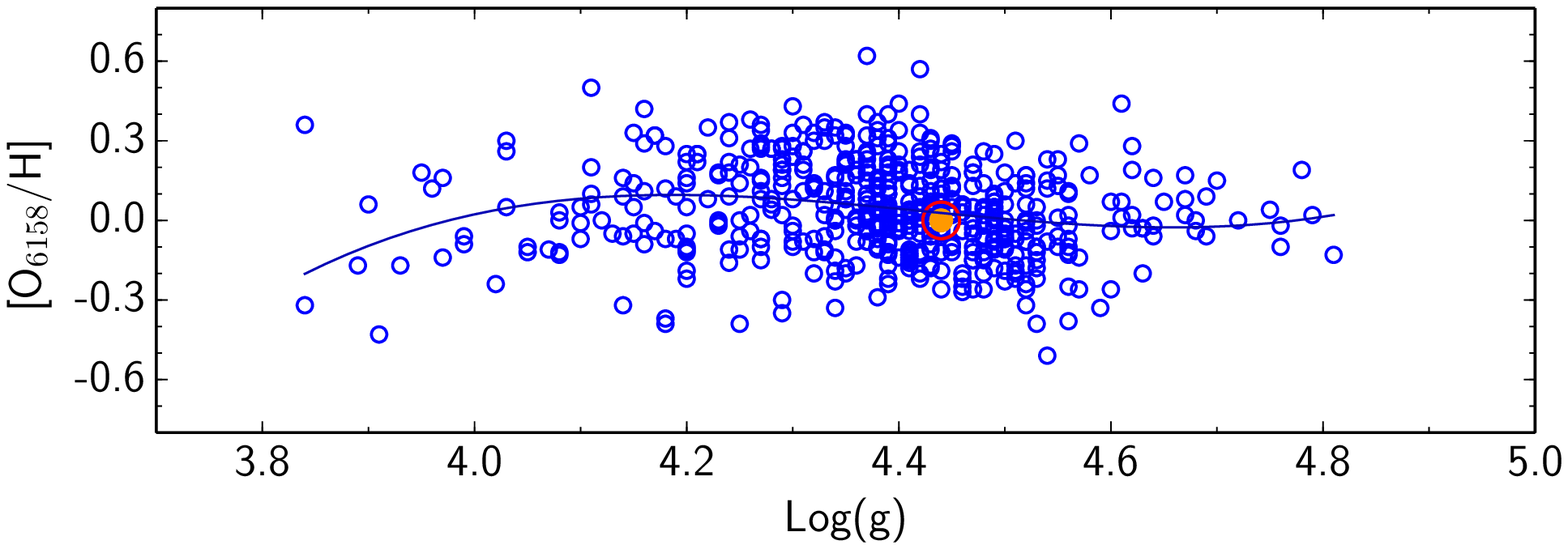}}
	\end{minipage} 
\caption{Systematic trends of [O/H] for different indicators with temperature and surface gravity. The solid line is the cubic best fit.}
\label{TrendsParams}
\end{figure*}

\begin{table*}[ht]
\caption{Oxygen abundances from the OI 6158$\AA$ line}
\label{O6158Results}
\centering
\begin{tabular}{lcccccccc}
\hline\hline \\[-8pt]
Star & T$_{eff}$ & log(\textit{g}) & $\xi_{t}$ & [Fe/H] & EW$_{OI}$ & $\delta$EW$_{OI}$ & [O/H]$_{OI}$ & Gal.Pop. \\
 & (K) & (cm s$^{-2}$) & (km s$^{-2}$) & & (m$\AA$) & (m$\AA$) & & \\[2pt]
\hline \\[-8pt]
Sun&5777&4.44&1.00& 0.00&3.7&0.3& 0.00&thin\\
HD70889&6051&4.49&1.13& 0.11&6.9&0.5& 0.10&thin\\
HD21161&5923&4.24&1.14& 0.09&7.5&2.8& 0.18&thin\\
HD37226&6178&4.16&1.61&-0.12&7.4&0.8&-0.01&thin\\
HD125881&6036&4.49&1.10& 0.06&5.8&0.6& 0.04&thin\\
HD117105&5889&4.41&1.13&-0.29&3.0&0.7&-0.12&thin\\
...&...&...&...&...&...&...&...&...\\[2pt]
\hline
\end{tabular}
\tablefoot{The full table is available at the CDS.}
\end{table*}

\begin{table*}[ht!]
\caption{Oxygen abundances from the [OI] 6300$\AA$ line}
\label{O6300Results}
\centering
\begin{tabular}{lcccccccccc}
\hline\hline \\[-8pt]
Star & T$_{eff}$ & log(\textit{g}) & $\xi_{t}$ & [Fe/H] & [Ni/H]\tablefootmark{a} & EW$_{Ni}$ & EW$_{[OI]}$ & $\delta$EW$_{[OI]}$ & [O/H]$_{[OI]}$ & Gal.Pop. \\
 & (K) & (cm s$^{-2}$) & (km s$^{-2}$) & & & (m$\AA$) & (m$\AA$) & (m$\AA$) & & \\[2pt] 
\hline \\[-8pt]
Sun&5777&4.44&1.00& 0.00& 0.00&2.0&3.4&0.3& 0.00&thin\\
HD48611&5337&4.51&0.69&-0.36&-0.37&1.2&1.9&0.8&-0.42&thin\\
HD70889&6051&4.49&1.13& 0.11& 0.09&1.8&3.3&0.4& 0.10&thin\\
HD21161&5923&4.24&1.14& 0.09& 0.11&2.2&5.9&2.9& 0.23&thin\\
HD130322&5365&4.37&0.90&-0.02&-0.03&2.5&5.1&1.0& 0.09&thin\\
HD222422&5475&4.46&0.73&-0.12&-0.15&1.8&2.8&0.8&-0.16&thin\\
...&...&...&...&...&...&...&...&...&...&...\\[2pt]
\hline
\end{tabular}
\tablefoot{
\tablefoottext{a}{[Ni/H] from \citet{Adibekyan12}.} The full table is available at the CDS.
}
\end{table*}

In Fig.~\ref{TrendsParams} we plot oxygen abundances against stellar parameters. Such plots reveal systematic effects that could influence our measurements. We note a slight dependence of [O$_{6300}$/H] on T$_{eff}$ at temperatures less than 5600K, for which lower [O/H] abundances are found. Although restricted to the low temperature end, a trend with opposite sign can also be noticeable for [O$_{6158}$/H]. Nevertheless, if we only consider stars with temperatures larger than $\sim$5400 K (e.g.\ when the 6158$\AA$ line is strong enough for reliable measurements) we will find no temperature dependence for the 6158$\AA$ line. In general, given the high abundance scatter observed in these graphs, the small dependence on T$_{eff}$ for cool stars can be neglected. No gravity effect was found.

%%%%%%%%%%%%%%%%%%%%%%%%%%%%%%%%%%%%%%%%%%%%%%%%%%%
% UNCERTAINTIES
%%%%%%%%%%%%%%%%%%%%%%%%%%%%%%%%%%%%%%%%%%%%%%%%%%%
\subsection{Uncertainties}
There are several sources of random error involved in the process of measuring chemical abundances. These are related either to the measurement of the EW or to the calculation of the abundances. It is not easy to define the contribution of each error and account analytically for them. In the present work we have considered three different contributions to the uncertainties in our measurements.

\begin{table}[ht]
 \caption[]{\label{ParamGroups}Average stellar parameters for the three sub-samples of different T$_{eff}$.}
 \centering 
\begin{tabular}{ccccc}
 \hline \hline \\[-8pt]
   &T$_{eff}$ &log(\textit{g}) & [Fe/H] &$\xi_{t}$ \\
   &  (K) & & & (km s$^{-1}$) \\[2pt]
  \hline \\[-8pt]
  Low  T$_{eff}$ & 5292 & 4.35 & -0.16 & 0.73\\
  Solar T$_{eff}$ & 5774 & 4.40 & -0.12 & 1.00\\
  High T$_{eff}$ & 6340 & 4.50 & -0.10 & 1.59\\[2pt]
\hline
\end{tabular}
\end{table}

First, we studied the sensitivity of the abundances to the stellar parameters: effective temperature T$_{eff}$, metallicity [Fe/H], surface gravity log(\textit{g}) and microturbulence $\xi_{t}$. Variations of these parameters introduce small changes in the abundances which, regardless of the parameter we vary, also depend on the T$_{eff}$ of the star \citep{Adibekyan12}. Therefore, we will carry on our analysis with three different sub-samples based on the T$_{eff}$: T $<$ 5377, 5377 $\leq$ T $\leq$ 6177 and T $>$ 6177. The middle group comprises the stars with T$_{eff} = $T$_{\odot}\pm$ 400 K. We expect that the sensitivity to the stellar parameters behave similar within each range of temperature. The average values of the stellar parameters for each group are shown in Table~\ref{ParamGroups}. These parameters together with the average oxygen abundance on each group are used to compute theoretical EWs of oxygen lines using the \textit{ewfind} driver of MOOG. Keeping the EW constant, we then calculate abundances by varying, one by one, the stellar parameters by  an amount equal to their uncertainty (Table~\ref{ParamErrors}). Among all the values, we select as the final error the largest deviation from the original oxygen value. 6158$\AA$ appears to be more sensitive to T$_{eff}$ while 6300$\AA$ larger variations with log(\textit{g}). 

\begin{table*}[ht]
 \caption[]{\label{ParamErrors}Abundance sensitivities to the stellar parameters of both oxygen indicators for the three sub-samples of different T$_{eff}$.}
 \centering 
\begin{tabular}{llcccccccc}
 \hline \hline \\[-8pt]
    & & \multicolumn{2}{c}{$\Delta$T$_{eff}$} & \multicolumn{2}{c}{$\Delta$log(\textit{g})} & \multicolumn{2}{c}{$\Delta$[Fe/H]} & \multicolumn{2}{c}{$\Delta\xi_{t}$}\\
    & & +30~K & -30~K & +0.06~dex & -0.06~dex & +0.03~dex & -0.03~dex & +0.08~kms$^{-1}$ & -0.08~kms$^{-1}$ \\ [2pt]
  \hline \\[-8pt]
  \multirow{2}{*}{Low T$_{eff}$} & \ \ $\Delta$~OI~6158 & -0.03 & +0.04 & +0.02 & -0.01 & \ \ 0.00 & +0.01 & \ \ 0.00 & +0.01 \\
  & \ \ $\Delta$~[OI]6300 & +0.01 & \ \ 0.00 & +0.03 & -0.02 & +0.02 & -0.01 & \ \ 0.00 & \ \ 0.00  \\
  \multirow{2}{*}{Solar T$_{eff}$} & \ \ $\Delta$~OI~6158 & -0.02 & +0.03 & +0.02 & -0.01 & \ \ 0.00 & +0.01 & \ \ 0.00 & \ \ 0.00 \\
  & \ \ $\Delta$~[OI]6300 & +0.01 & \ \ 0.00 & +0.03 & -0.02 & +0.01 & -0.01 & \ \ 0.00 & \ \ 0.00 \\
  \multirow{2}{*}{High T$_{eff}$} & \ \ $\Delta$~OI~6158 & -0.02 & +0.02 & +0.02 & -0.02 & \ \ 0.00 & +0.01 & \ \ 0.00 & \ \ 0.00 \\
  & \ \ $\Delta$~[OI]6300 & +0.01 & -0.01 & +0.02 & -0.02 & +0.01 & -0.01 & \ \ 0.00 & \ \ 0.00 \\[2pt]
\hline
\end{tabular}
\end{table*}

Another source of uncertainty is the statistical photometric error due to the noise in each pixel. We can evaluate the partial derivative of the EW with respect to the flux \citep{Bohlin83,Cayrel88} to obtain an analytical solution to this contribution:
%\noindent
   \begin{displaymath}
   EW=\sum\left(\Delta\lambda\frac{C_{i}-I_{i}}{C_{i}}\right),
   \delta EW=\sqrt{\sum_{i=1}^N\left(\frac{\partial EW}{\partial I_{i}}\right)^2}=
   2.45\epsilon\sqrt{\sigma\Delta\lambda}
   \end{displaymath}
Where $C_{i}$ is the continuum flux level, $I_{i}$ is the flux of each measurement, $\Delta\lambda$ represents the wavelength increment per pixel (~0.01$\AA$ for HARPS spectra), $\epsilon$ is the r.m.s.\ relative photometric accuracy in the continuum and $\sigma$ is the standard width over which we sum six times assuming a gaussian profile. We determine the standard width theoretically as the quadratic sum of different broadenings: instrumental (wavelength divided by the spectral resolution of the instrument), macroturbulence based on the relation provided by \citet{Valenti05}, microturbulence, rotation (using an average value of 3~kms$^{-1}$)  and thermal broadening. The latter is calculated with \citep{Gray92}:
   \noindent
   \begin{displaymath}
   \Delta\lambda_{D} = \frac{\lambda}{c}\sqrt{\frac{2kT}{m}} = 4.301\cdot10^{-7}\lambda\sqrt{\frac{T}{\mu}} 
   \end{displaymath}
Where $k$ is Boltzmann's constant, T represents the effective temperature and $\mu$ is the atomic weight in atomic mass units. 

The last and the greatest contribution to the uncertainties is related to the measurement of the EW and its sensitivity to the continuum placement. Although this error can reach a large percentage of the EW, it is commonly ignored by many authors, resulting in significantly underestimated uncertainties in the abundances. In order to quantify this contribution, one should calculate the difference between the areas of the gaussian fits by placing the continuum in different positions. Given the high S/N of our spectra, an therefore the low uncertainty in the rms, the possible continuum placements will yield small variations between the respective Gaussian depths. Thus, the error in EW can be approximated as the area of the rectangle between the best and the $\pm\epsilon$ continuum placements, where $\epsilon$ is the inverse of the S/N per resolution element. The width at the continuum level is again equal to 6$\sigma$. All together yields an error of $\pm6\sigma\epsilon$.

\begin{figure}[]
	\resizebox{\hsize}{!}{\includegraphics{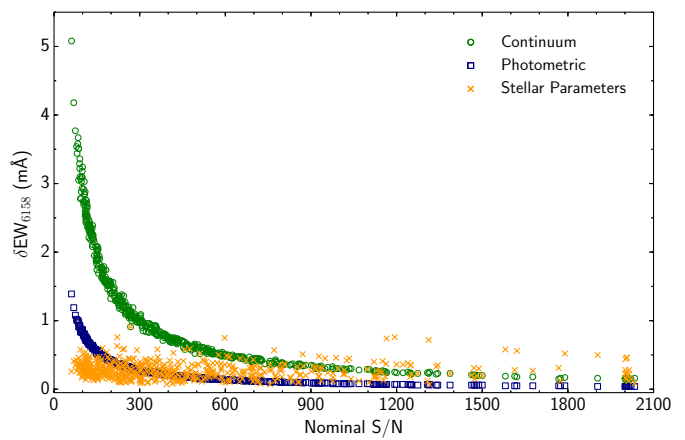}}
\caption{Contribution to the total equivalent width error of OI6158$\AA$ of each source of uncertainty.}
\label{eEW}
\end{figure}

Figure~\ref{eEW} shows significance of different contributions to the total uncertainty of the EW, which is similar for both oxygen lines. We have converted the error in abundance due to uncertainties in stellar parameters to the corresponding EW error using the \textit{ewfind} driver of MOOG synthesis code. Each contribution to the EW is finally added quadratically to get the one-$\sigma$ uncertainty. Because of the non-linear dependence of the abundance on the EW, asymmetric errors will arise when the EW error is propagated to get abundance uncertainties. For further details on the determination of the abundance uncertainty we refer the reader to the appendix.

%%%%%%%%%%%%%%%%%%%%%%%%%%%%%%%%%%%%%%%%%%%%%%%%%%%
% COMPARISON INDICATORS
%%%%%%%%%%%%%%%%%%%%%%%%%%%%%%%%%%%%%%%%%%%%%%%%%%%
\section{Comparison between indicators}
\label{SectionComp}

Abundances from both the OI6158$\AA$ and [OI]6300$\AA$ lines were measured in 447 stars from the initial sample of 762. This large number of stars covers a wide range of temperatures and metallicities, which allows us to carry out a comprehensive comparison of the performance of these spectral lines as oxygen abundance indicators. In this section we study the agreement between chemical abundances obtained from both spectral lines, and we also investigate their reliability and usability related to the stellar parameters.

\begin{figure}[ht]
	\centering
	\resizebox{0.68\hsize}{!}{\includegraphics{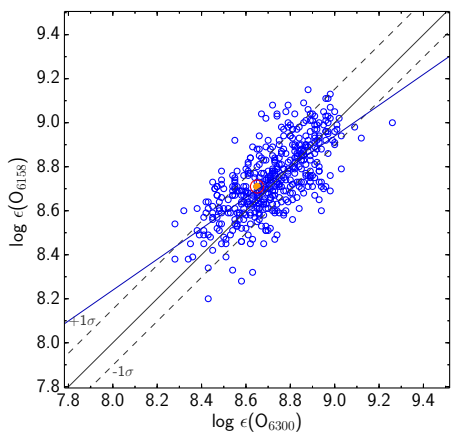}}
	\caption{Comparison of the absolute oxygen abundances from OI6158$\AA$ with those obtained from [OI]6300 $\AA$ lines. Solid lines represent the diagonal and the linear fit to the data. Dashed lines enclose 68\% of the data around the mean (0.024~dex).}
	\label{OH6300vsOH6158}
\end{figure}

In section \ref{Analysis} we have already noted that the absolute oxygen abundances from the two lines in the Sun disagree, since log~$\epsilon$(O$_{6158}$) is 0.06~dex higher than log~$\epsilon$(O$_{6300}$). The comparison for all the stars with common measurements is shown in Figure~\ref{OH6300vsOH6158}. The OI6158$\AA$ line provides higher abundances in stars with [O/H] similar or lower than the Sun, while [OI]6300$\AA$ yields higher values for oxygen-rich stars. This small trend could be due to overestimation of the equivalent width when the lines are small. However, the trend is not significant as it falls within the standard deviation of $\pm$1$\sigma = $~0.127~dex. On average, absolute oxygen abundances derived from the 6158$\AA$ spectral line are 0.024~dex higher than [OI]6300$\AA$. This value is considerably smaller than the typical uncertainties of the oxygen abundances. No conclusion can be drawn, except that on average both spectral lines yield compatible values of oxygen abundance within the uncertainties.

\begin{figure}[ht]
	\centering
	\resizebox{1.05\hsize}{!}{\includegraphics{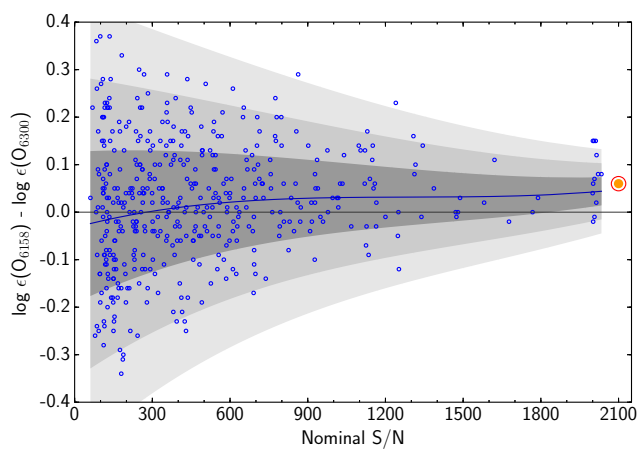}}
	\caption{Difference between absolute oxygen abundances derived from the OI6158$\AA$ and [OI]6300$\AA$ lines against nominal signal-to-noise ratio. Weighted moving average (boxcar average) of data points is shown as a solid line, together with the 1, 2, and 3 sigma dispersion regions in different grey shades. Dashed lines represent the simple moving average of the abundance errors for each line. In the top panel the Sun has been shifted towards lower S/N for plotting purposes. The lower panel shows the width of the $\pm$1$\sigma$ region for the data average and the uncertainties.}
	\label{SNRvsDifLines}
\end{figure}

As we move towards stars with better quality spectra, i.e. higher S/N, the abundance uncertainties decrease. As expected, our measurements become more precise and we achieve a smaller dispersion in the difference between oxygen indicators. Specifically, while the standard deviation for the whole sample was 0.127~dex, at high S/N it drops to 0.072~dex. However, for stars with S/N$>$1000 the average difference in oxygen abundances is 0.051~dex, which is compatible with previously measured differences in the Sun \citep{Caffau08}. These deviations increase as we consider higher S/N ratios, indicating a possible discordance between both indicators. The agreement between abundance indicators as a function of the S/N is presented in figure~\ref{SNRvsDifLines}. We have performed a weighted moving average (500 steps, step size 50 and weigthed with 1/$\sigma^2$) that has been smoothed by a factor of 0.4. The 1,2 and 3 standard deviations levels from the mean, which are represented by shaded areas in the figure, clearly show how the dispersion decreases at better S/N, while the mean is shifted towards higher values of O6158$\AA$. This result may suggest that the 0.06~dex disagreement found in the Sun is not unique, and may have some physical explanation.

The  different sensitivities of the two oxygen lines to stellar parameters allows us to define windows where each of these lines can be used more or less easily. The equivalent width of both spectral lines is smaller than 8m$\AA$ at T$_{eff}>$5800K. However, because of the high excitation energy of 6158$\AA$, this line becomes significantly stronger in hotter stars. Specifically, the EW of OI6158$\AA$ is always larger than the EW of [OI]6300$\AA$ in all the stars with T$_{eff}>$6200K.

\begin{figure}[ht]
	\centering
	\resizebox{\hsize}{!}{\includegraphics{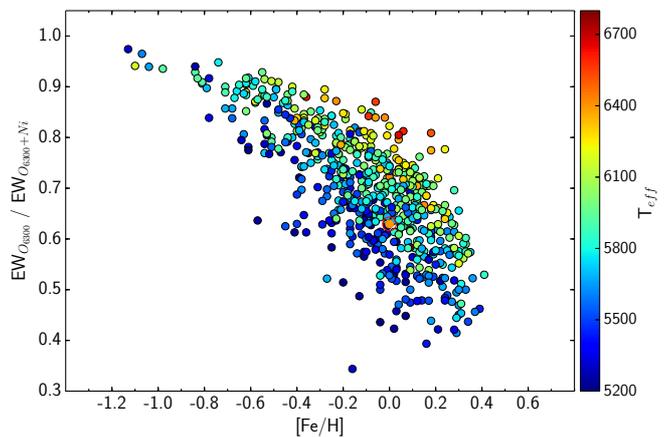}}
	\caption{Fraction of oxygen in the equivalent width of the 6300 $\AA$ feature as a function of metallicity. Dependence on the effective temperature is shown in colours.}
	\label{PercentEW6300}
\end{figure}

It is also important to note the uncertainties that affect each line. Because of the density of spectral features in the range of 6158$\AA$, the largest source of error in the equivalent width measurement is the continuum placement. Therefore, it is more complicated to measure this line in cool stars, where in addition to many blends, the line itself is very weak.  The 6300$\AA$ feature can be considered as problematic in solar-type stars for two important reasons. The first is related to the presence of numerous telluric lines often severely blended with the oxygen feature. Even more  serious is the problem with the Ni absorption line (see above). \citet{Johansson03} provided a log(\textit{gf}) value with 14\% uncertainty. Our calculations show that the Ni line can account for up to 60\% of the EW of the 6300$\AA$ absorption feature in cool, metal-rich stars (see Fig~\ref{PercentEW6300}). Thus, the abundance of Ni and its atomic parameters plays a key role in this topic. As an illustration, the error in the Ni log(\textit{gf}) is introducing an uncertainty in the oxygen EW measurement of HD142 equal to 3.1$^{+0.2}_{-0.1}$~m$\AA$, which is translated into [O/H]=0.2$^{+0.03}_{-0.00}$~dex. In stars with medium to low S/N this contribution will be smaller than the two main sources of uncertainty considered in this work. Even though, the larger scatter found in figure~\ref{OH} with 6300$\AA$ could be partially explained by the uncertainty in the Ni log(\textit{gf}), in combination with the possible presence of undetected telluric lines slightly above the level of the noise.
   
The main and the most important aspect of the 6158$\AA$ line is that it does not suffer from known blends. It is as strong as the forbidden line in solar analogues and stronger than 6300$\AA$ at temperatures higher than 6200K. Our results suggest that the 6158$\AA$ line can be used as a reliable indicator of oxygen abundance 
in solar-type stars. However, we should bear in mind  that it may suffer non-negligible non-LTE effects in hot stars \citep{Sitnova13}.

%%%%%%%%%%%%%%%%%%%%%%%%%%%%%%%%%%%%%%%%%%%%%%%%%%
% Galactic trends
%%%%%%%%%%%%%%%%%%%%%%%%%%%%%%%%%%%%%%%%%%%%%%%%%%%
\section{Oxygen trends in the thin and thick disks}
\label{GalTrends}

\subsection{Observed dispersion of [O/H]}

The abundance scatter seen in the Galactic trends of many chemical elements is very small. \citet{Adibekyan12} carried out a homogeneous study and found small abundance dispersion in such elements as Si or Ti. The abundances presented here have been derived from the same spectra and stellar parameters. Thus, using the same database and techniques as \citet{Adibekyan12} we would expect to get equally precise and trustworthy results for other elements. However, those elements have many more reliable spectral lines for abundance analysis, and therefore is not surprising that we found a wider scatter in the case of oxygen (fig.~\ref{OH}). Even so, scatter found in the present work for [O$_{6158}$/H] is comparable to that of Mg, which was derived using three spectral lines \citep{Adibekyan12}. \citet{Petigura11} suggested an astrophysical origin to explain the dispersion found in oxygen trends.

\begin{figure}[ht]
	\centering
	\resizebox{\hsize}{!}{\includegraphics{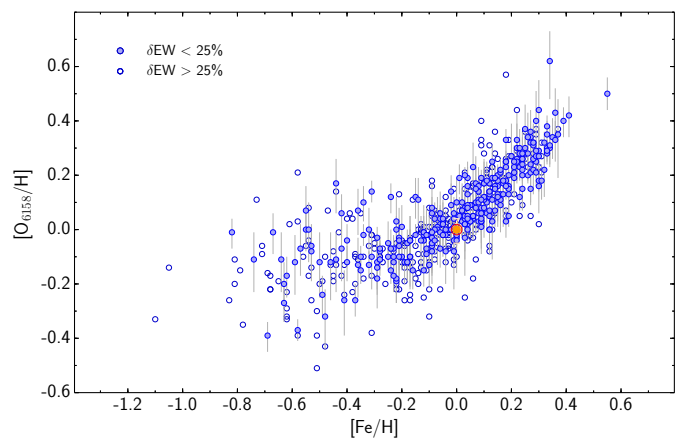}}
	\caption{[O/H] vs.\ [Fe/H] for OI 6158 $\AA$. Filled (open) circles represent data for which EW measurements have an uncertainty under (above) 25\%. Absolute error bars are over-plotted for our best measurements.}
	\label{OH_err}
\end{figure}

\begin{figure}[ht]
	\centering
	\resizebox{\hsize}{!}{\includegraphics{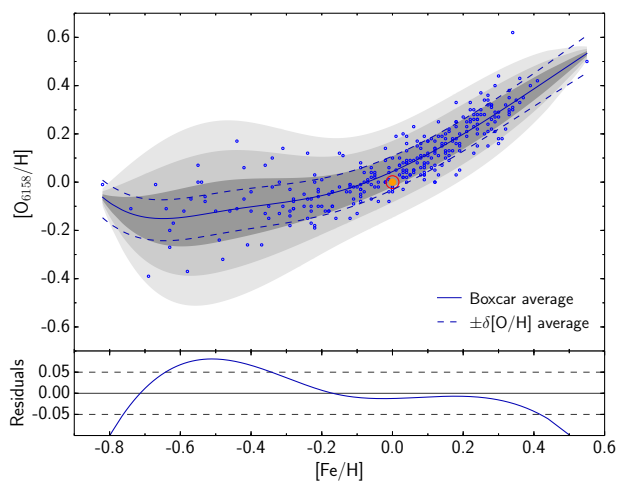}}
	\caption{\textit{Upper panel:} [O/H] vs.\ [Fe/H] for OI 6158 $\AA$. Only stars with EW relative error below 25\% are considered. Solid line represent the average oxygen abundance. The 1,2 and 3 sigma levels of dispersion around the mean are presented as shaded areas. Dashed line shows the average uncertainties. For plotting purposes the average uncertainties are shown around the mean abundance. \textit{Lower panel:} Difference between the $\pm1\sigma$ width of the scatter around the mean and the $\pm1\sigma$ width of the uncertainties.}
	\label{OH_boxcar}
\end{figure}

We consider a sub-sample of 296 stars, 55\% of the sample, for which we have errors smaller than 25\% in EW in order to verify whether the observed abundance scatter of oxygen is due to the lack of precise measurements. We also restricted the analysis to the OI 6158 line, since we found in this study that it is more trustworthy. We see in Fig.~\ref{OH_err} that the scatter is still very high in this `most precise' sub-sample at [Fe/H] $<$ $-$0.2~dex. On the other hand, at higher metallicities the scatter is not remarkably reduced by selecting the best measurements, and only outliers are removed. Error bars are about the size of the dispersion and it is therefore difficult to draw any conclusion about the origin of the scatter. 

The observed scatter at a given metallicity  is expected to have two contributions: a real star-to-star scatter, and also a dispersion due to the uncertainties in our measurements. We have performed a moving average (250 steps with 0.03~dex width) to derive both the average [O/H] with their respective scatter, and the mean $\pm\delta$[O/H] as a function of metallicity. The goal of this test is to find out which of the two contributions to the scatter dominates our sample. Results of the test are depicted in fig.~\ref{OH_boxcar}. We are aware that this test is not valid at the edges of the metallicity range due to the undersampling of the bins, which makes unreliable the provided mean values.

% Two column figure 
\begin{figure*}[ht]
   \centering
	\begin{minipage}[b]{0.49\linewidth}
		\resizebox{\hsize}{!}{\includegraphics{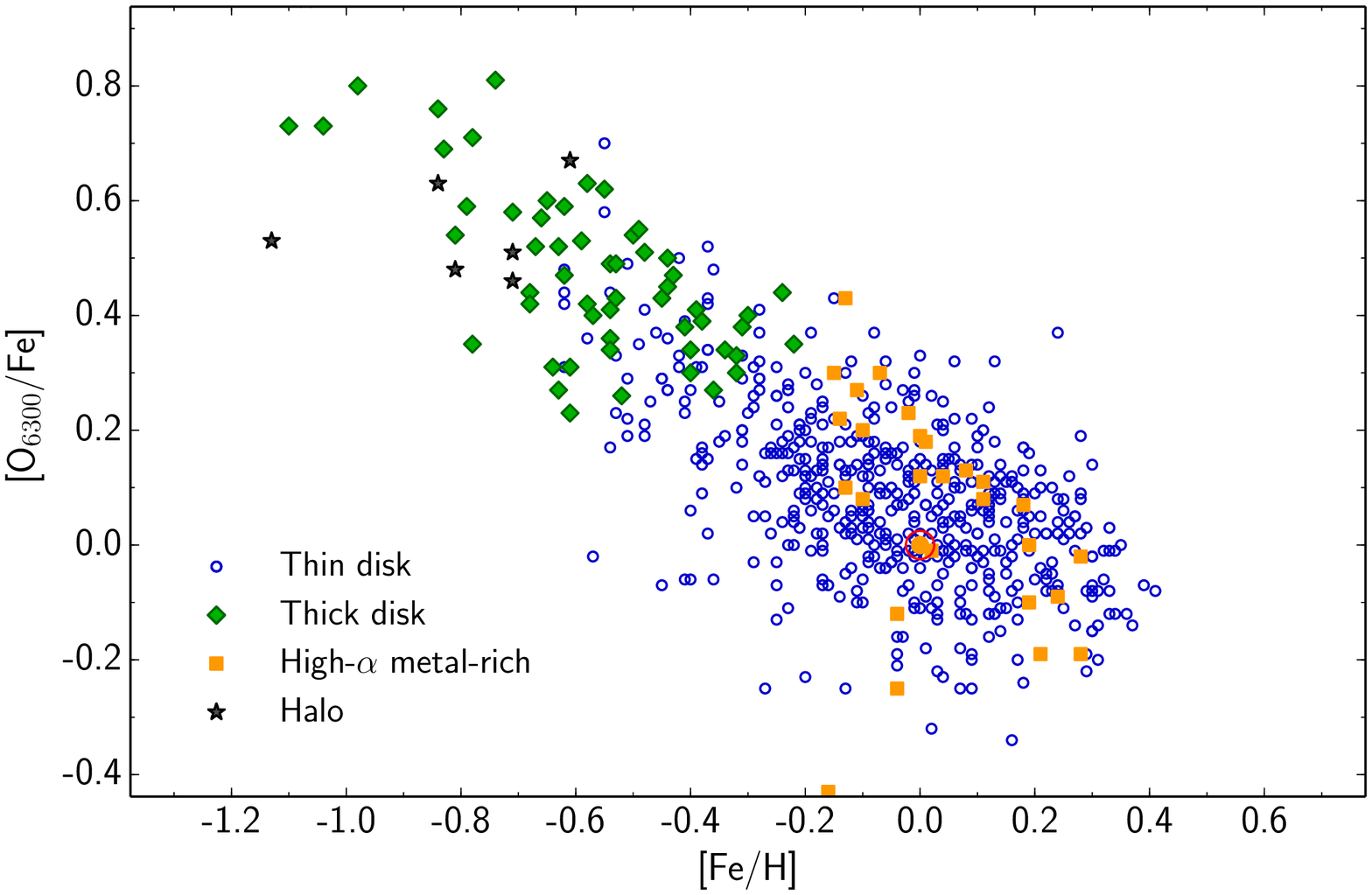}}
	\end{minipage} 
	\begin{minipage}[b]{0.49\linewidth}
		\resizebox{\hsize}{!}{\includegraphics{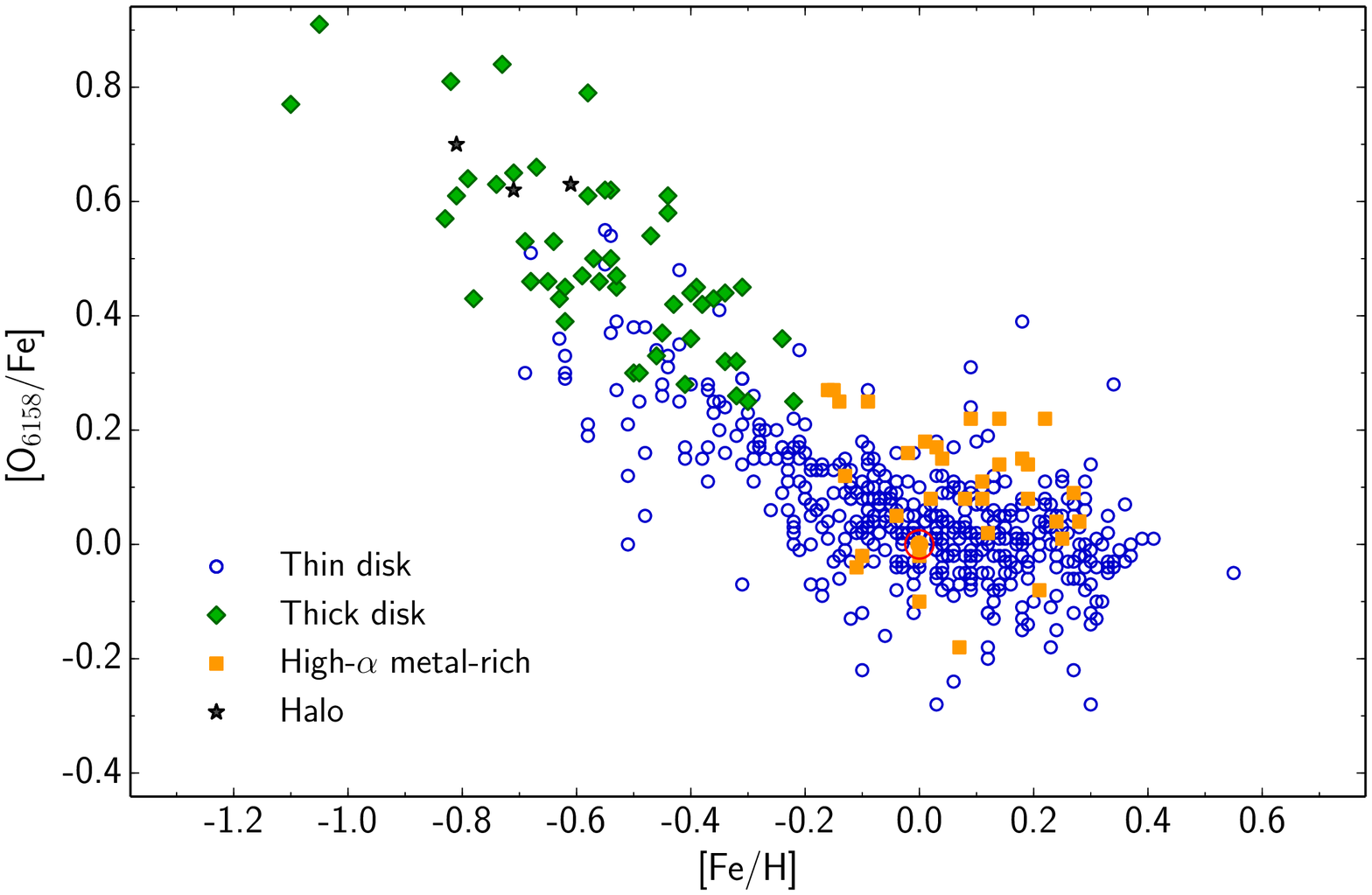}}
	\end{minipage} 
\caption{[O/Fe] vs.\ [Fe/H] for [OI] 6300 $\AA$ and OI 6158 $\AA$. Stars from different populations are separated according to their chemical composition.}
\label{OFeGal}
\end{figure*}

In order to evaluate the relative importance of the contributions to the scatter, in the lower panel of the figure we show the difference between the widths of the $\pm$1$\sigma$ regions. A negative residual would imply an overestimation of our uncertainties, while a positive residual could be due to either a real scatter or to an underestimation of the errors. Since the uncertainties have been derived with the same method for all stars, and there is no dependency on metallicity, differences in residuals as a function of metallicity are expected to be caused by real star-to-star dispersion. At high metallicities the observed scatter and the uncertainties are comparable, and do not allow us to detect the possible presence of an astrophysical scatter. However, the positive residual found at [Fe/H]$<$-0.2~dex reveals a real star-to-star scatter that cannot be explained by uncertainties. The results of this test are not surprising, and they are in perfect agreement with the fact that at [Fe/H]$<$-0.2~dex there is a mixture of stars from two different populations of the Galactic disk (see below for details). However, this expected results confirm the effectiveness of this method to show up the presence of a real scatter. In addition, the satisfactory agreement between uncertainties and scatter at high metallicities support the validity of our uncertainty calculations. Given that the test was performed using a sub-sample of the best measurements, this last result should consequently be interpreted with caution. A similar test with the whole sample would show a slight overestimation of the uncertainties (negative residuals), due to the exponential growth of the errors as we move towards lower S/N. Therefore, we conclude that although our uncertainty calculations are overestimated at very low S/N, they appear to satisfactory account for the observed scatter at medium to high S/N.

\subsection{Thin and thick disk separation}

It is believed that $\alpha$ elements, such as oxygen, are produced in massive stars that enrich the interstellar medium when they explode as type II supernova. By contrast, iron peak elements are mostly synthesized in SNe Ia explosions \citep[e.g.][]{Thielemann02}. The oxygen over iron ratio can provide hints on the formation of the Galactic disk and its chemical history, but the observed Galactic trends are diffused as a result of the mixture of stars from two different populations in the studied metallicity regime.

\begin{figure}[ht]
   \centering
	\resizebox{\hsize}{!}{\includegraphics{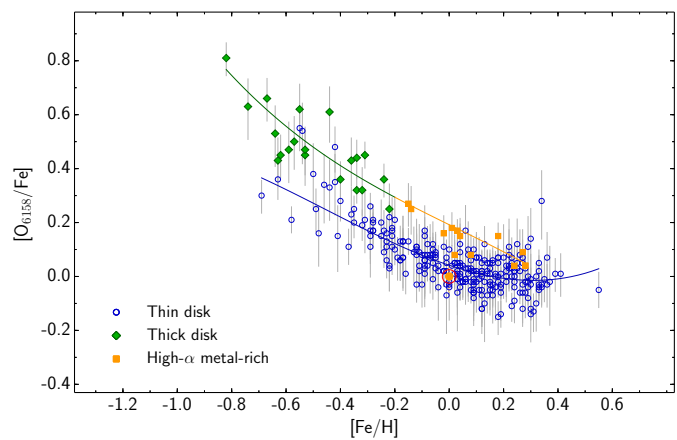}}
	\caption{[O/Fe] vs.\ [Fe/H] for the stars with $\delta$ EW $<$ 25\%. Stars from different populations are separated according to their chemical composition. Solid lines represent the fitted trend, weighted with the uncertainties, for each population (thick and h$\alpha$mr stars are fitted as a single population). Absolute error bars are over-plotted.}
\label{O6158Gal}
\end{figure}

The chemical distinction between thin and thick disks was first noticed by \citet{Fuhrmann98}  and further investigated by \citet{Prochaska00}, among others. We have classified the stars in our sample between thin disk, high-$\alpha$ metal-poor (thick disk) and high-$\alpha$ metal-rich stars according to chemical criteria described in \citet{Adibekyan11}. These authors show that disk populations are clearly separated in terms of [$\alpha$/Fe], where `$\alpha$' refers to the average abundance of Mg, Si, and Ti. The ratio of oxygen over iron for the different disk populations is presented in Figure~\ref {OFeGal}. There is no obvious difference seen between the thick and the thin populations from [OI]6300, which can be attributed to the larger errors and scatter. We see a clear overabundance of oxygen in thick disk stars once the OI6158$\AA$ line is considered. Moreover, using the sub-sample with the most precise abundance values (Fig.~\ref{O6158Gal}), we can check that none of the thick disk stars falls below the trend for the thin disk, although few thin disk stars have [O/Fe] higher than expected for this population. In addition, both trends appear to be remarkably separated at lower metallicities. Thus, we can confirm with our data that oxygen, just like the other $\alpha$~elements, is enhanced in the thick disk stars.

The high-$\alpha$ metal-rich population (h$\alpha$mr) was presented by \citet{Adibekyan11,Adibekyan13} as a new population due to the gap in metallicity found between the latter and the thick disk. In addition, a gap in the distribution of [$\alpha$/Fe] at 0.17 was also noticed. The present work is based on the same sample as \citet{Adibekyan11}, and therefore our data also show a gap in metallicity. The separation in the [O/Fe] distribution is not as clear as in [$\alpha$/Fe] because both populations are mixed around [O/Fe]$\sim$0.25, probably caused by the larger errors in the determination of oxygen abundances compared to other $\alpha$~elements. Our most precise sample show that all the stars from the h$\alpha$mr population fall above the average trend for thin disk (Fig.\ref{O6158Gal}), and thus we can confirm that h$\alpha$mr stars are enhanced in [$\alpha$/Fe] (where $\alpha$ refers here to oxygen). Recent studies by \citet{Bensby14} and \citet{Helmi14} also confirmed the alpha enhancement of a metal-rich population, nevertheless, they did not observe a gap in metallicity between them and the thick disk stars. \citet{Bensby14} argued that the high-$\alpha$ metal-rich family might be a metal rich tail of the thick disk. With this new oxygen abundance determination we cannot draw conclusions about this population, i.e. whether is distinct or just a continuation of the thick disk towards higher metallicities.

\begin{figure}[ht]
	\centering
	\resizebox{\hsize}{!}{\includegraphics{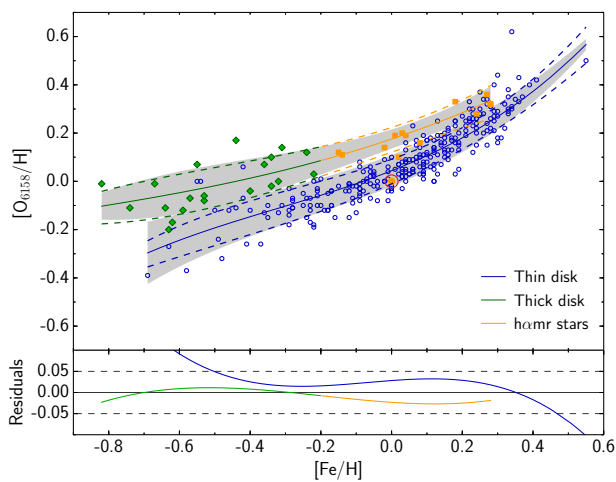}}
	\caption{\textit{Upper panel:} [O/H] vs.\ [Fe/H] for OI 6158 $\AA$. Only stars with EW relative error below 25\% are considered. Stars are sepparated in different disk populations according to a chemical criteria. Solid line represent the average oxygen abundance. The 1$\sigma$ level of dispersion around the mean is presented as a shaded area. Dashed line shows the average uncertainties. For plotting purposes the average uncertainties are shown around the mean abundance. \textit{Lower panel:} Difference between the $\pm1\sigma$ width of the scatter around the mean and the $\pm1\sigma$ width of the uncertainties.}
	\label{OH_boxcar_pop}
\end{figure}

Once the disk populations are separated, we can revisit the issue about the origin of the scatter in the observed [O/H] trends. We repeat the test performed for the best sample, but this time independently for the thin and thick (including h$\alpha$mr) disk populations. The moving average for the oxygen abundances and $\pm\delta$[O/H] uncertainties are shown in figure~\ref{OH_boxcar_pop}, together with the residuals of the $\pm$1$\sigma$ width regions. Given the small residuals found between uncertainties and scatter for thick and h$\alpha$mr population, and taking into account the reduced number of stars that we have in this group, no conclusion can be drawn about the existence of a real star-to-star dispersion. In contrast, thin disk stars seem to show a positive residual at -0.2$<$[Fe/H]$<$0.3, which reaches 0.03dex at [Fe/H]$\sim$0.1. This residual may reveal a real astrophysical scatter, since uncertainties are unlikely to be underestimated. On the other hand, it can be checked that the thick and thin disk populations appear to be separated at a 1$\sigma$ level, thus confirming the enrichment of $\alpha$-elements in the thick disk. In addition, the enhancement of the h$\alpha$mr population is clearly revealed.

\subsection{Comparision with previous studies}

Several studies have addressed the behaviour of oxygen over iron in the thin and thick disk populations. In general, all of them agree that thick disk stars show an enhancement of $\alpha$ elements, being this caused by different ratios of SNII/SNIa in both disks. Some of them found a shallow decline in the [O/Fe] trend at high [Fe/H] for thin disk stars \citep[e.g.][]{Bensby04, Ramirez07, Takeda05, Petigura11, Bensby14}, indicating that in the thin disk both SN Ia and SN II contribute at a steady rate to the enrichment of the interstellar medium. In contrast, \citet{Ramirez13} and \citet{Nissen02, Nissen14} results suggest a flattening of [O/Fe] at super-solar [Fe/H]. Our data also suggest that flattening in the [O/Fe] ratio may appear at solar metallicities (Fig.~\ref{OFeGal}). The ``knee'' is inappreciable in the case of [OI], and thus, for the moment we cannot rule out that a change of slope below [Fe/H] $\sim$ 0 is real. This apparent lack of agreement between both oxygen indicators at high [Fe/H] can be attributed either to the scatter in the [OI]6300 data or because some   blend may appear at high metallicities near the OI6158$\AA$ line, leading to an overestimation of the EW. However, although further work needs to be done to study the linelist near 6158$\AA$ spectral line, neither the FWHM or the symmetry of the line seem to suggest the presence of an unknown blend.

% Two column figure 
\begin{figure*}[ht]
   \centering
	\begin{minipage}[b]{0.33\linewidth}
		\resizebox{\hsize}{!}{\includegraphics{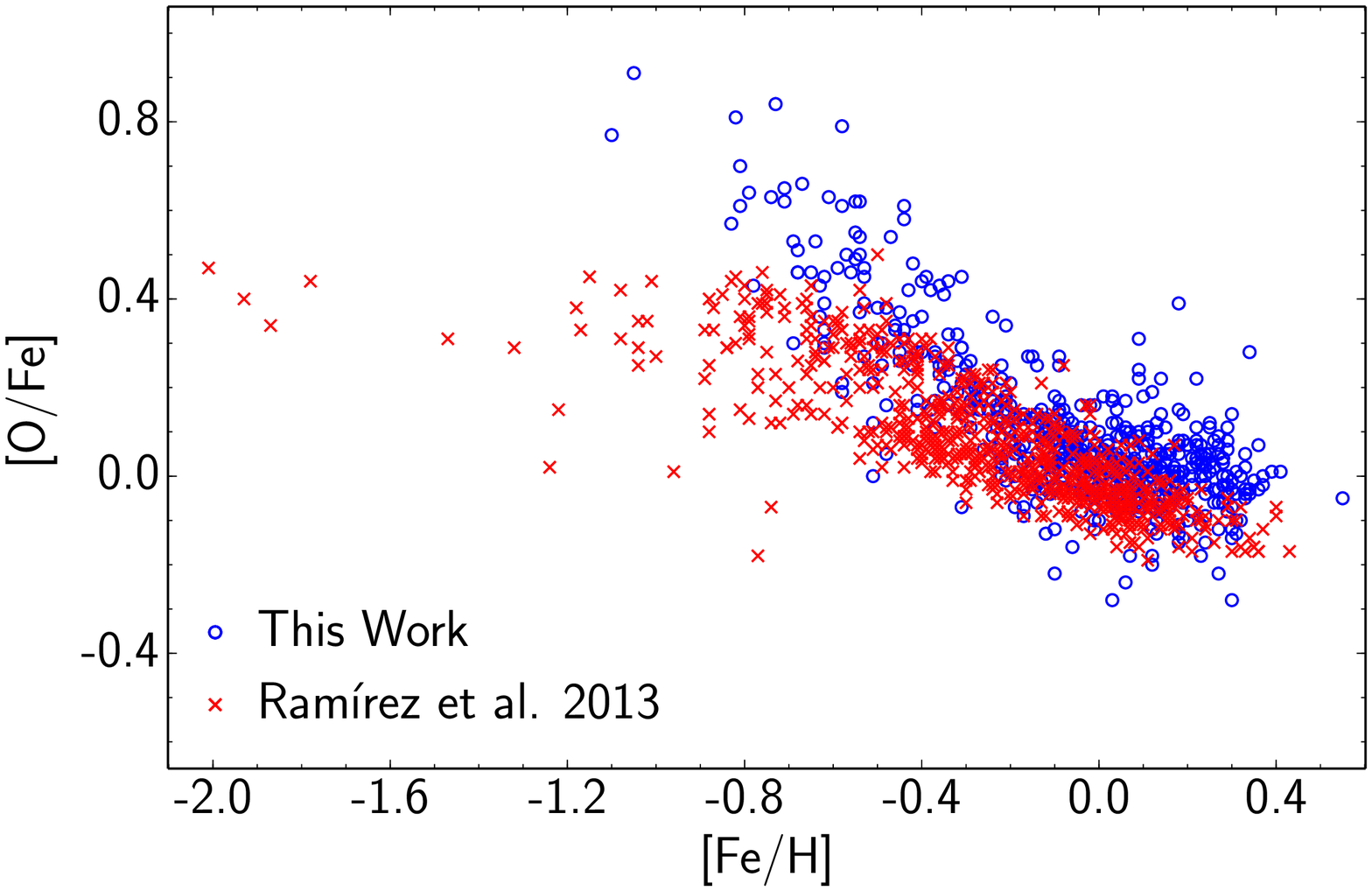}}
	\end{minipage} 
	\begin{minipage}[b]{0.33\linewidth}
		\resizebox{\hsize}{!}{\includegraphics{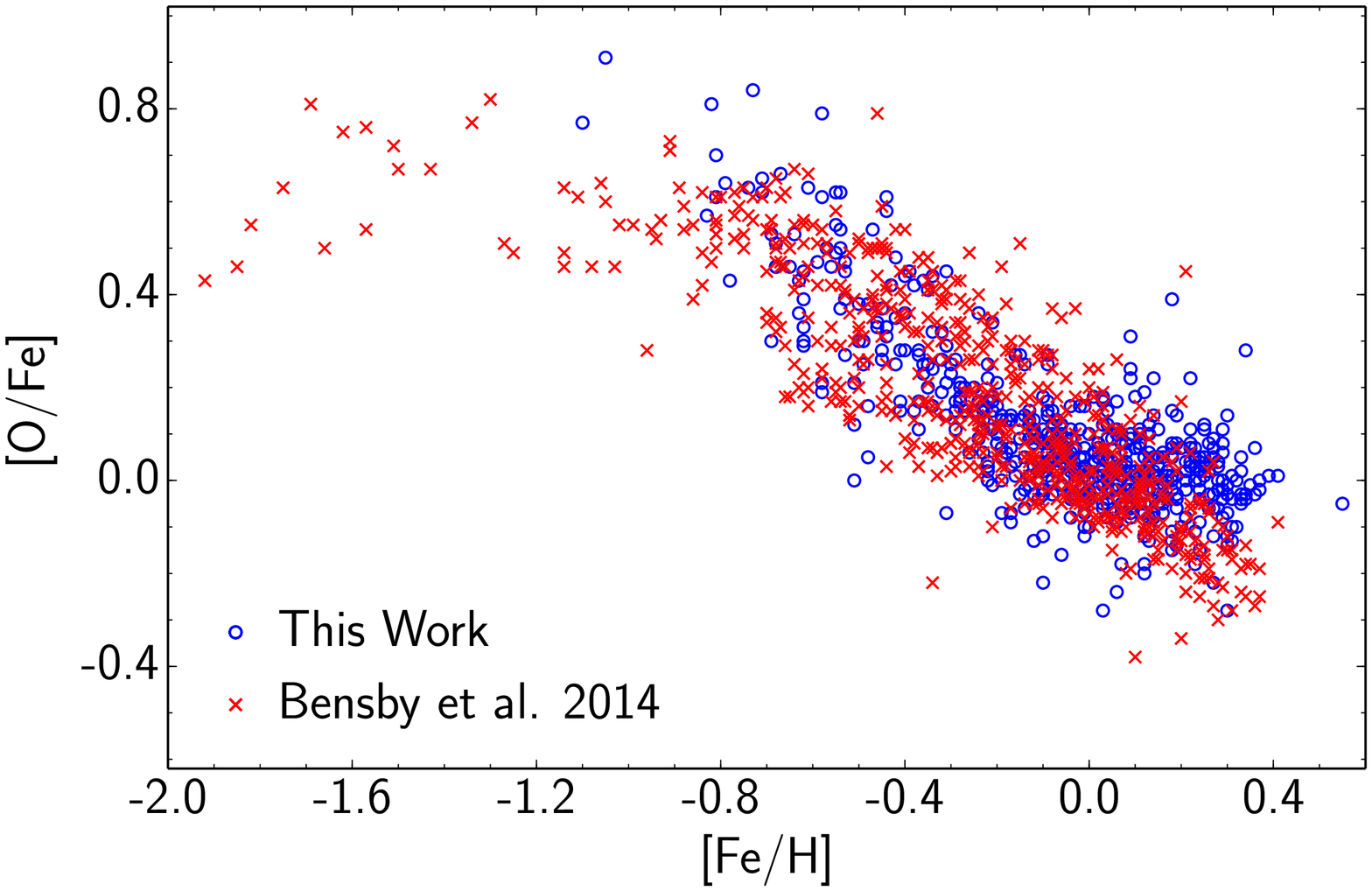}}
	\end{minipage} 
	\begin{minipage}[b]{0.33\linewidth}
		\resizebox{\hsize}{!}{\includegraphics{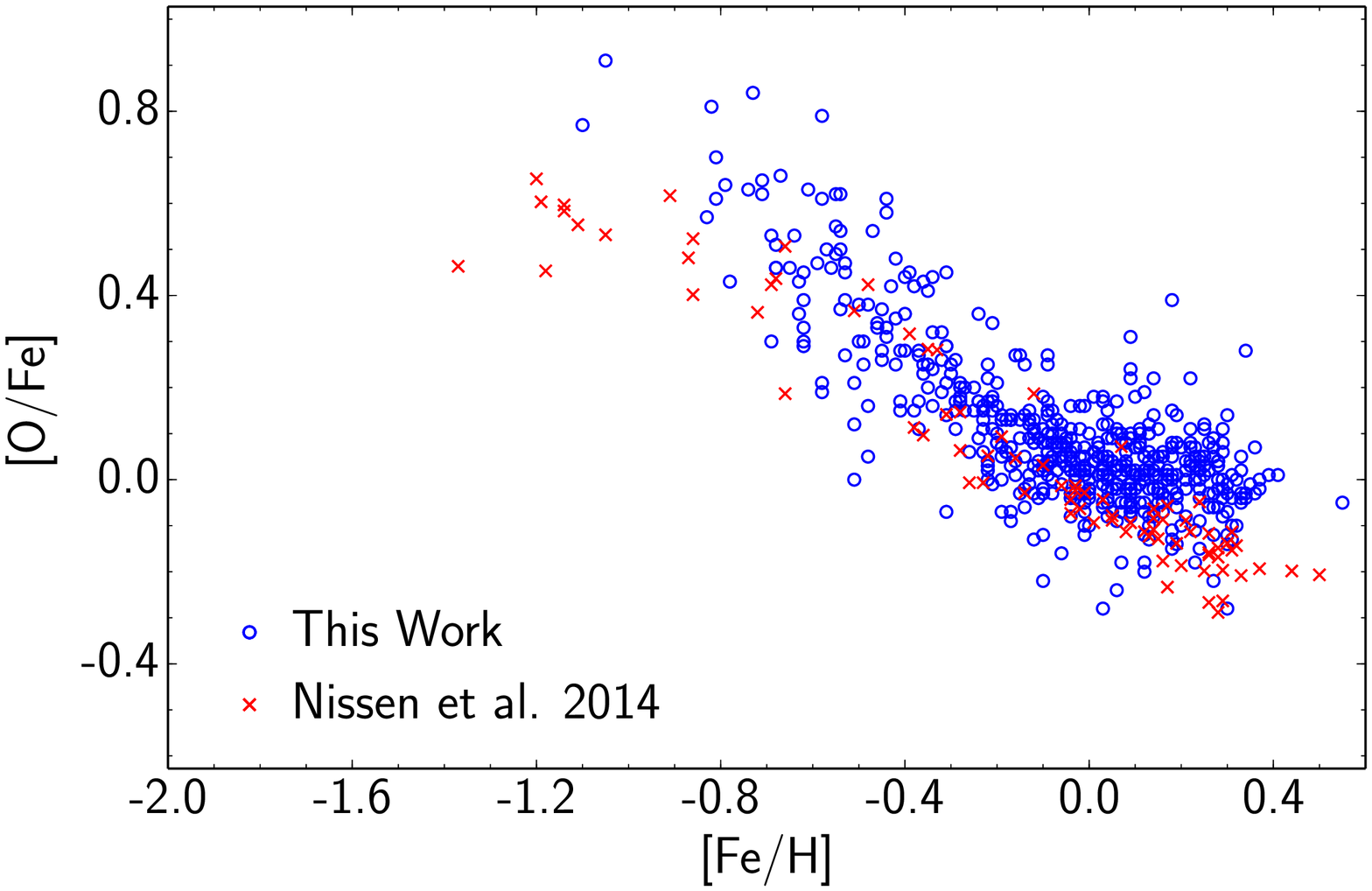}}
	\end{minipage}
\caption{Comparison of our  [O$_{6158}$/Fe] vs.\ [Fe/H] trends to those derived by \citet{Ramirez13}, \citet{Bensby14} and \citet{Nissen14}. For comparison purposes, abundance values from \citet{Ramirez13} and \citet{Nissen14} have been rescaled to the same solar reference (log~$\epsilon$(O)$_\odot$=8.71). \citet{Bensby14} performed a normalization relative to their solar values on a line-by-line basis. However, individual line abundances are not provided, which makes not possible to rescale to the same solar reference.}
\label{Comparison}
\end{figure*}

In order to study the behaviour of [O/Fe] towards the metal-poor regime in Fig.~\ref{Comparison} we compare our results of OI6158$\AA$ with those of \citet{Ramirez13}. These authors derive oxygen abundances from the infrared triplet and correct them for non-LTE effects, but neglect inelastic collisions with hydrogen. We also make a comparison with the results of \citet{Bensby14} and \citet{Nissen14}, both of them based on a combination of samples for which oxygen was measured using [OI]6300 and the OI IR triplet, and only the OI IR triplet, respectively. \citet{Nissen14} corrected oxygen abundances for departures from LTE accounting for H collisions, as described by the classical Drawin formula \citep{Drawin68}, and scaled by S$_{H}$=0.85. \citet{Bensby14} used for this purpose an empirical correction derived by comparing abundances of 60 stars from [OI]6300 and OI774 indicators. 

\begin{figure}[ht]
	\resizebox{\hsize}{!}{\includegraphics{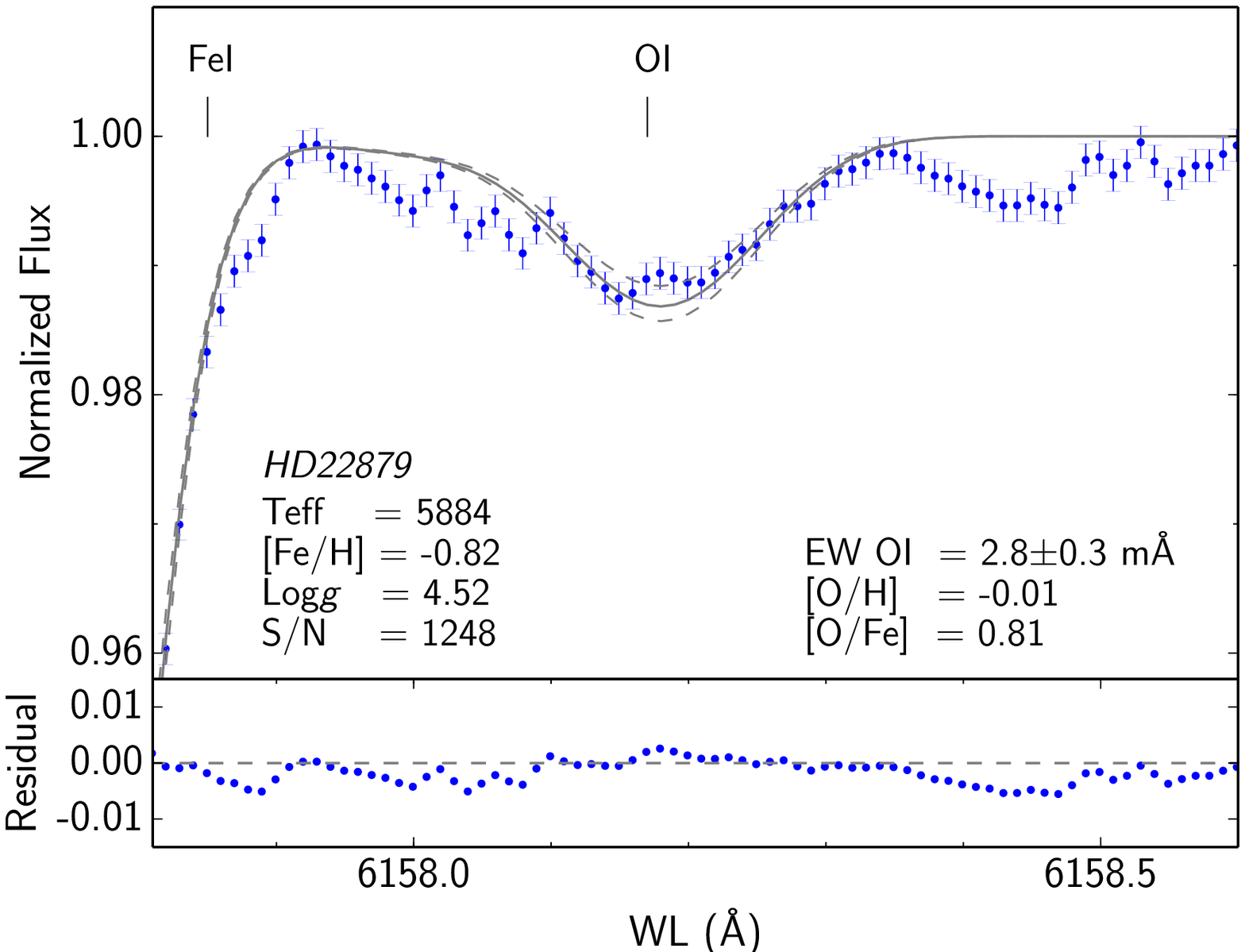}}
	\resizebox{\hsize}{!}{\includegraphics{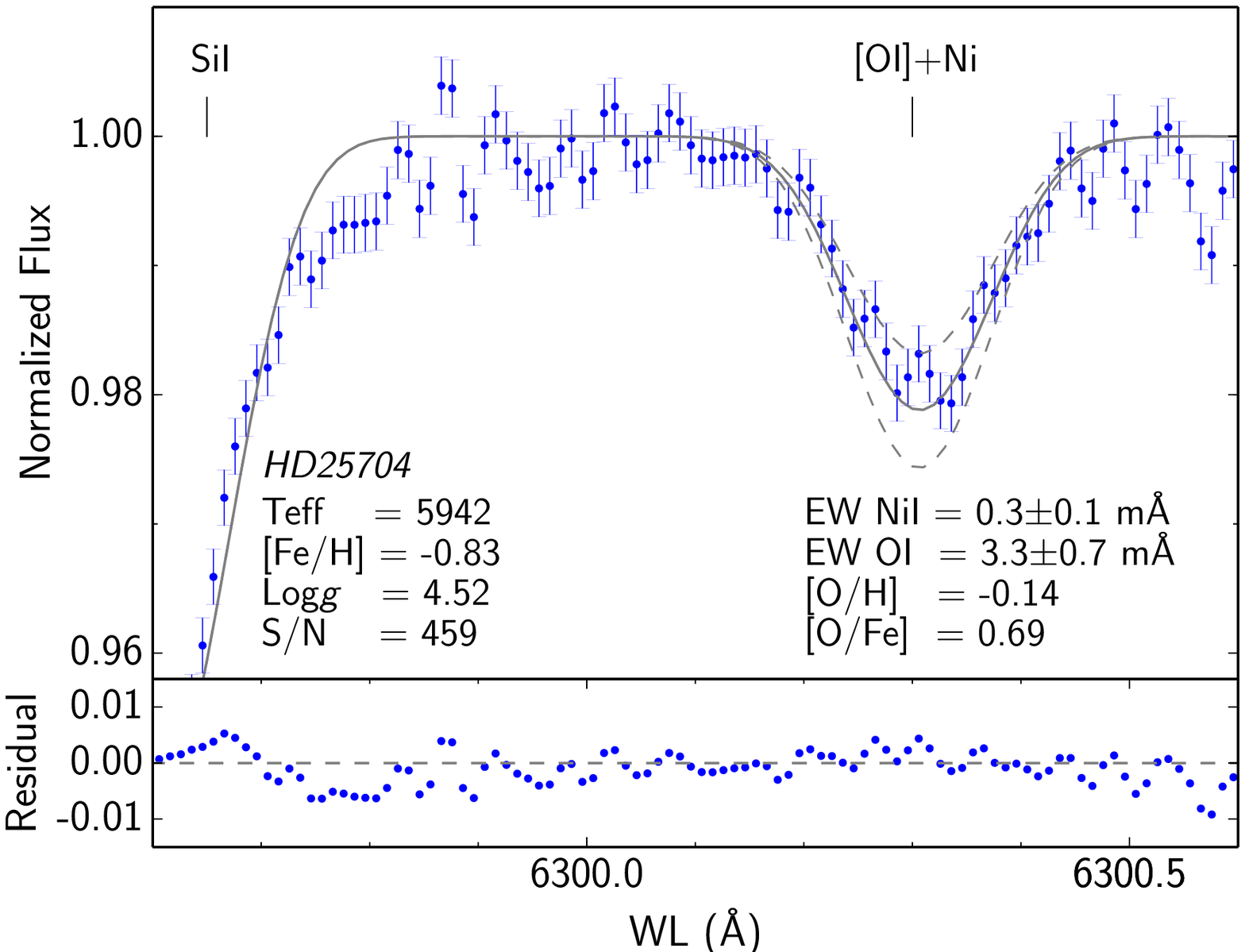}}
\caption{Spectral synthesis of the oxygen line at 6158$\AA$ in HD\,22879 (upper panel) and [OI]6300 in HD\,25704 (lower panel). The observed spectra (dots) with the 1$\sigma$ photometric noise (rms) is compared with the synthetic spectra, computed for the corresponding [Fe/H] and [O/H] abundances obtained from our EW measurements (solid line). Dashed lines represent the synthetic spectra for the $\pm1\sigma$ error abundances, which are $\delta$[Fe/H]$=\pm0.03$~dex, $\delta[O_{6158}/H]=^{+0.06}_{-0.07}$~dex and  $\delta[O_{6300}/H]=^{+0.13}_{-0.18}$~dex (propagated from the provided EW error). Also shown are the stellar parameters, nominal S/N, abundances, EW and residual intensities (observed - predicted).}
\label{HighOFe}
\end{figure}

While our results mostly agree with those of \citet{Bensby14} and \citet{Nissen14}, we find clear disagreement for the metal-poor stars of \citet{Ramirez13}. Fig.~\ref{Comparison} shows at [Fe/H]$<$-0.5 how the [O/Fe] rise becomes steeper and we obtain much larger values for [O/Fe] than \citet{Ramirez13}. On the other hand, despite the similar slope, our results reach higher [O/Fe] ratios than \citet{Nissen14} and \citet{Bensby14} at [Fe/H]$\sim$-0.8~dex. Although the most metal-poor stars in our sample have weak OI6158$\AA$ oxygen lines, with EW between 2 and 4 m$\AA$, we can safely rule out the possibility that our values are overestimated because of the poor continuum measurements and/or other sources of errors. To illustrate this point, in Fig.~\ref{HighOFe} we show one of the targets in our sample HD\,22879 with one of the highest [O/Fe]=0.81 and [Fe/H]=$-$0.82. The combined spectrum of this star has S/N=1000 and our precise measurements of OI6158$\AA$ EW, which perfectly match the synthetic spectra, clearly rule out an EW weaker than 2.5~m$\AA$. Thus, our best value for this star is  [O/Fe]=0.81. These high oxygen over iron ratios are also found with [OI] measurements, as an example HD25704 is a star with [Fe/H]=-0.83 and [O/Fe]=0.69~dex. Finally, it is noteworthy to mention that the observed trend towards higher values of [O$_{6158}$/H] at low temperatures (Fig.~\ref{TrendsParams}) do not affect the observed [O/Fe] trend since none of the high [O/Fe] stars have T$_{eff}<$5600K.

Given the quality of the observations and measurements presented in these studies, the simplest way to explain the discrepancy with other works rely on the different non-LTE corrections applied to the abundance calculations. Differences with \citet{Ramirez13} arise because the inelastic collisions with H atoms were neglected in their OI IR triplet calculations, which would have shifted the trend towards the LTE results. Actually, calculations of non-LTE by \citet{Fabbian09} taking into account inelastic collisions with neutral oxygen conclude that non-LTE effects presented by these authors were overestimated by $\sim$0.10~dex at [Fe/H]$\sim$-1.5~dex. On the other hand, the disagreement found between [OI]6300 and OI IR triplet in the Sun lead us to question the non-LTE correction applied by \citet{Bensby14}. Although this correction may be a good first approach to the problem, it would be useful to determine it with more accuracy. Given the current knowledge about departures from the LTE, and the weight of hydrogen collision to bring the system closer to the LTE, \citet{Nissen14} approach appears to be the most reasonable. 

The effects of departures from LTE for 6158$\AA$ transitions have been  widely studied in the Sun. Unfortunately, there is not much literature about non-LTE corrections at lower [Fe/H] for this spectral line. As a high-excitation line, it is formed in deep layers of the atmospheres, where the source function and line opacity effects work in opposite directions. Thus, the combined effect results in very small non-LTE effects in the Sun $\sim$-0.03~dex \citep{Caffau08}. However, generally speaking, departures from LTE are more severe towards higher T$_{eff}$ and lower log(\textit{g}) and [Fe/H]. Therefore, one would expect larger non-LTE corrections in the low-[Fe/H] tail of our [O/Fe] distribution. Temperature is not expected to increase the effect of non-LTE in our stars, for which effective temperatures are between 5600 to 6200K at low metallicity. \citet{Sitnova13} studied non-LTE effects for oxygen permitted transitions in different stars and for a grid of parameters, taking into account hydrogen collisions (S$_{H}$=1). According to their work, non-LTE corrections are not expected to be larger than -0.05~dex in our sample.

%%%%%%%%%%%%%%%%%%%%%%%%%%%%%%%%%%%%%%%%%%%%%%%%%%
% DISCUSSION
%%%%%%%%%%%%%%%%%%%%%%%%%%%%%%%%%%%%%%%%%%%%%%%%%%
\section{Discussion}

In this article we show that the [O/Fe] ratio rises steeply with decreasing metallicity, and reaches 0.8 around [Fe/H]=-1.  Our data are not extended below [Fe/H]= -1 and it is not clear if this ratio will stay constant  or keep rising in halo stars down to [Fe/H]$<$-3.  Obviously both oxygen lines employed in this work will disappear in halo dwarfs since their EWs will be less than 1m$\AA$. Thus,  there is no chance to extend this study to F type metal-poor halo dwarfs. Finally, we note that despite considerable observational effort the trend of the [O/Fe] ratio in the halo is still uncertain. However, most of the studies \citep{Israelian01,  Takeda03, Fulbright03} suggest that main-sequence dwarfs provide more reliable and consistent abundances than giants. 

A steep rise of the [O/Fe] ratio and high values above 0.7 found in the present analysis can hardly be  explained in the framework of classical models of galactic chemical evolution. We consider, for example, the models proposed by \citet{Chiappini01}. This model considers two main accretion episodes for the formation of the Galaxy, the first forming the halo and bulge in a short timescale and the second one forming the thin disk.  They propose a short halo formation timescale of about 0.8~Gyr. At a certain epoch (coinciding with the halo-disk transition) core-collapse SNe, responsible for the production of O, stopped exploding while Fe, produced by the long-living SNe Ia, continued to be produced. Star formation must have ceased for a period that cannot be longer than 1~Gyr. The two-infall model fits the observed metallicity distribution of the G-dwarfs by assuming a long timescale for the thin-disk formation. These models have been calculated assuming a constant plateau  at [Fe/H]$<$-1 for all $\alpha$-elements including oxygen, for which they propose [O/Fe]=0.5 at [Fe/H]=-1. These models predict a light increase in [O/Fe] towards lower [Fe/H] reaching [O/Fe]$\sim$0.4 at [Fe/H]=-1. However, some authors \citep[e.g.][]{Boesgaard11} have already reported that $\alpha$-elements do not show a plateau at [Fe/H]$<$-1. Moreover,  the steep rise of [O/Fe] does not necessarily imply the same behaviour for other $\alpha$-elements \citep[e.g.][]{Ramaty00}.  The different trend for oxygen compared to the other $\alpha$-elements at [Fe/ H]$>$0 can also be explained by the fact that oxygen is only  produced in Type II SN. The levelling out of the [$\alpha$/Fe] trend for other $\alpha$ elements is clearly expected since they also have small contributions from Type Ia SN. Metallicity dependent oxygen yields will make the overall picture more complete and may provide additional clues to explain these observations \citep{Prantzos94}.

Any model for the thick disk origin should account for several observational facts. The first one is the old age of thick disk stars and the temporal gap (1-2~Gyr) with respect to the oldest thin disk stars. Then, the tight relationship found for several abundance ratios in the thick disk. And finally, the variation of the abundance ratios with metallicity. It will be an interesting challenge for these models to explain observations of $\alpha$-elements from the same HARPS database \citep{Neves09, Adibekyan12} as well as [O/Fe] trend ratio presented here since they come from the same homogenous sample. 

Different models have been proposed to explain observational characteristics of the thick disk. It is possible that thick disks have been created through accretion of galaxy satellites \citep{Meza05, Abadi03}, where thick-disk stars therefore have an extragalactic origin. It has also been proposed that thick disks were born thick at high redshift from the internal gravitational instabilities in gas-rich clumpy disks \citep{Bournaud09, Forbes12}. Another possibility is that thick disks are created through the heating of pre-existing thin disks with the help of mergers \citep{Quinn93, Villalobos08}, whose rate decreases with decreasing redshift. One could propose that a massive merger will create very intensive star formation  at [Fe/H]=-1 and produce stars with very high [O/Fe] ratios.

The thick and thin disks have formed on different timescales. Our data shows that the average ages of thin and thick disks are 4.73~Gyr  and 9.20~Gyr, respectively. Details about age determination are provided by \citet{Delgado14}. In fact, all our stars with [O/Fe]$>$0.6 belong to the thick disk and are older than 6~Gyr. The Galactic thin disk has not had such an intense star formation history as the thick disk. The shallow decline in the [O/Fe] is caused by continuous star formation with no fast initial enrichment from Type II SN. 

In the recent past there has been a growing conviction that radial migration can be responsible for the formation of thick disks by bringing out high-velocity-dispersion stars from the inner disk and the bulge \citep[e.g.][]{Schonrich09}. While by now it has been established that because of conservation of vertical action migration does not contribute to disk thickening \citep{Minchev12, Martig14, Vera-Ciro14}, radial migration could bring stars with high [O/Fe] from inner galaxy to solar neighbourhood \citep{Minchev13}.

The stellar evolution calculations for massive stars \citep{Woosley95, Thielemann96} predict that the ejected mass of (O/Fe) by a massive star is an increasing function of the stellar mass. Thus, an increase of the slope of the IMF during the halo/thick disk transition phase could possibly explain our observations.  A very steep linear increase of the [O/Fe] in  thick disk stars could also be due either to a lower contribution of Fe from Type II SNe or to a metallicity dependent yield of oxygen (or a combination of these two effects). The change of metallicity on oxygen yield has a different impact depending on the mass range of stars considered \citep{Maeder92}.

Perhaps the easiest way to explain high [O/Fe] ratios in metal-poor stars is by modifying the so called mass-cut parameter in the models of explosive supernova. Present explosive nucleosynthesis calculations for Type II SN are based on induced supernova explosions by either depositing thermal energy or invoking a piston with a given kinetic energy. Induced calculations (lacking self consistency) utilize the constraint of requiring ejected $^{56}$Ni masses from the innermost explosive Si burning layers in agreement with supernova light curves being powered by the decay chain $^{56}$Ni-$^{56}$Co-$^{56}$Fe. This can also serve as guidance to the supernova mechanism with mass cuts based on $^{56}$Ni in the ejecta. It means that the position of a mass cut can define the amount of Fe ejected from the supernova  as well as the [O/Fe] ratio in the ejecta \citep{Woosley95, Thielemann96}.

%%%%%%%%%%%%%%%%%%%%%%%%%%%%%%%%%%%%%%%%%%%%%%%%%%
% Conclusion
%%%%%%%%%%%%%%%%%%%%%%%%%%%%%%%%%%%%%%%%%%%%%%%%%%%
\section{Summary and conclusions}
\label{Conclusion}

We present a uniform study of oxygen abundances in a large sample of F- and G-type dwarfs from the HARPS database. A homogeneous set of the atmospheric parameters was adopted, and two independent analyses for different indicators were performed. We provide the first accurate and homogeneous comparison of the OI6158$\AA$ and [OI]6300$\AA$ lines of oxygen.

Oxygen is one of the most controversial elements owing to the problems related to inconsistent abundances derived from different spectral lines. In this work we have investigated the behaviour of the lines at 6158$\AA$ and 6300$\AA$. Our measurements suggest that abundances derived from both oxygen lines are consistent within their respective error bars. However, we have found that for spectra with a very high S/N these values are no longer compatible within errors. Mainly, the abundance derived from  the OI6158$\AA$ line is 0.072~dex higher than that derived from the [OI]6300$\AA$ line. A similar difference has been previously reported for the Sun \citep{Caffau08}. Thus, our results indicate that the disagreement is not unique, and that there must be some physical explanation for this fact.

For F-type stars with temperatures higher than 6200 K, the use of OI6158$\AA$ is recommended. The equivalent widths of this line are larger than those of the [OI] 6300. Thus, the errors due to the placement of continuum are lower for OI 6158. This line can also be useful in the super-solar metallicity regime where the Ni 6300 line may contribute up to 50\% of the feature at 6300$\AA$. In addition, the uncertainty of  log(\textit{gf}) for Ni is non-negligible; therefore, errors in the measurement of [OI]6300$\AA$ could be significant. In contrast, at low temperatures and metallicity below solar values, OI6158$\AA$ becomes weak and yields more uncertain values of oxygen abundances. In any case, both lines can be used in solar analogues if S/N $>$ 300, allowing us to measure   equivalent widths as small as  2 m$\AA$. With this quality of data, the OI6158$\AA$ line also represents a good alternative when the [OI]6300$\AA$ line is severely blended with telluric features. It is always good to measure both lines whenever possible in order to check their consistency.

We found that the abundances derived from the OI6158$\AA$ line suffer from much smaller dispersion 
when considering the trends [O/H] and [O/Fe] vs.\ [Fe/H], in contrast with those derived from the [OI]6300$\AA$ line. The scatter found in the trends for a sub-sample of 252 high-precision measurements of OI6158$\AA$ is comparable to that of Mg, derived using three spectral lines \citep{Adibekyan12}. In addition, our results suggest that the [O/Fe] ratio reaches a constant value at super-solar metallicities, in contrast with some previous results that reported a shallow decline. A similar trend is seen in the results of \citet{Ramirez13} and \citet{Nissen14}.

Although there is no clear boundary between oxygen abundances in the thin and thick disks, the stars studied in this article show that some bifurcation might exist at [Fe/H] $<$ $-$0.3~dex, as found with the high-precision sub-sample of OI6158$\AA$. Our results also indicate a clear enhancement of oxygen abundances in the thick disk stars, as has been widely reported for other $\alpha$-elements.

%%%%%%%%%%%%%%%%%%%%%%%%%%%%%%%%%%%%%%%%%%%%%%%%%%
% Acknowledgements & Biblio
%%%%%%%%%%%%%%%%%%%%%%%%%%%%%%%%%%%%%%%%%%%%%%%%%%

\begin{acknowledgements}
We thank the anonymous referee for his/her thorough review and suggestions, which significantly contributed to improving the quality of the publication. We are also grateful to Dr. E. Pancino, Dr. C. Allende Prieto and Dr. A. del Pino for fruitful comments and discussion regarding the uncertainties calculations and statistical analysis. S.B. is grateful for financial support from the Spanish Ministry project MICINN AYA2011-29060. This work was also supported by the European Research Council/European Community under the FP7 through Starting Grant agreement number 239953. E.D.M., V.Zh.A. and S.G.S. are supported by grants SFRH/BPD/76606/2011,  SFRH/BPD/70574/2010 and SFRH/BPD/47611/2008 from FCT (Portugal), respectively. N.C.S. also acknowledges the support in the form of an Investigador FCT contract funded by FCT/MCTES (Portugal) and POPH/FSE (EC). This work has made use of TOPCAT\footnote{http://www.starlink.ac.uk/topcat/}.
\end{acknowledgements}

%-------------------------------------------------------------------

%\begin{thebibliography}{}
%\end{thebibliography}
\bibliographystyle{aa} % style aa.bst
\bibliography{oxygen} % your references Yourfile.bib

%\clearpage

%%%%%%%%%%%%%%%%%%%%%%%%%%%%%%%%%%%%%%%%%%%%%%%%%%
% Appendix
%%%%%%%%%%%%%%%%%%%%%%%%%%%%%%%%%%%%%%%%%%%%%%%%%%
\appendix
\section{Statistical approach to the determination of chemical abundances}
It is well known that the relation between the equivalent width and the abundance of a spectral line is non-linear. If we assume that the EW has a normal probability distribution, a non-linear transformation will yield an asymmetric abundance distribution. The most obvious consequence of this fact is the asymmetric abundance errors that one can find in any abundance analysis. However, these asymmetries are negligible for strong spectral lines. The present analysis is based on weak lines of oxygen, whose EW is smaller than 5 m\AA~in more than 50\% of the stars. Moreover, EW confidence intervals are not preserved under this non-linear transformation, which means that abundance errors derived from an LTE analysis does not correspond to the same  tolerance region. Aiming to provide statistically meaningful uncertainties, we carry out a detailed statistical error analysis.

\begin{figure}[]
	\resizebox{\hsize}{!}{\includegraphics{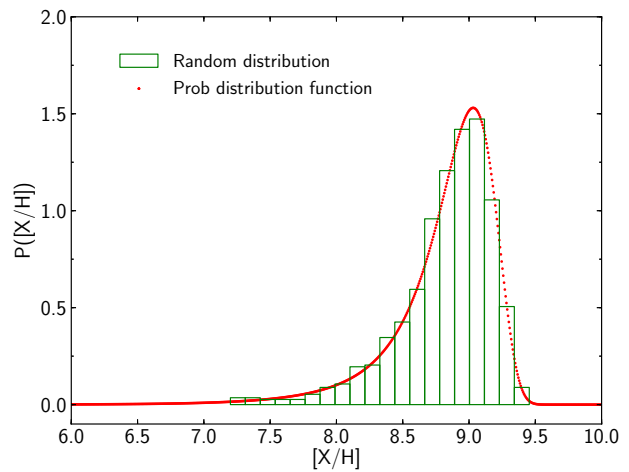}}
\caption{Abundance distribution probability for a large asymmetry case. Red dots represent the analytically derived function. Green bars represent the distribution of the abundance calculated with the curve of growth for a set of random EW values that follow a truncated gaussian distribution.}
\label{DistribProb}
\end{figure}

If we assume a normal probability distribution for the equivalent width, truncated at zero, we can obtain the abundance distribution as \citep{Barlow03}:
%\noindent
\begin{displaymath}
P([X/H]) = \frac{G_{T}(EW)}{\vert d[X/H]/dEW\vert}
\end{displaymath}
where G$_{T}$(EW) is the adopted distribution for the EW with mean equal to the measured EW and standard deviation calculated as the final 1$\sigma$ uncertainty of the EW. A simple relation between abundance and EW is provided by the curve of growth. Considering a single spectrum line, we can write:
%\noindent
\begin{displaymath}
\centering
mlog(EW) = [X/H] +c
\end{displaymath}
where m and c are constant coefficients. For each star we can find these coefficients by evaluating two values of EW, and deriving their abundances following the standard LTE analysis. The probability density function of the abundance will be:
\begin{displaymath}
P([X/H]) = \frac{ln10}{m\sigma\sqrt{2\pi}}10^{\frac{[X/H]-c}{m}}G_{T}\left( 10^{\frac{[X/H]-c}{m}}\right)
\end{displaymath}
We can test the correctness of this model by randomly generating a set of EW values following a normal distribution, truncated at zero. The assumed abundance distribution should be able to reproduce the distribution of the random EWs, transformed into abundance using the curve of growth. This test is shown in Figure \ref{DistribProb}.

% Two column figure 
\begin{figure*}[]
   \centering
	\begin{minipage}[b]{0.49\linewidth}
		\resizebox{\hsize}{!}{\includegraphics{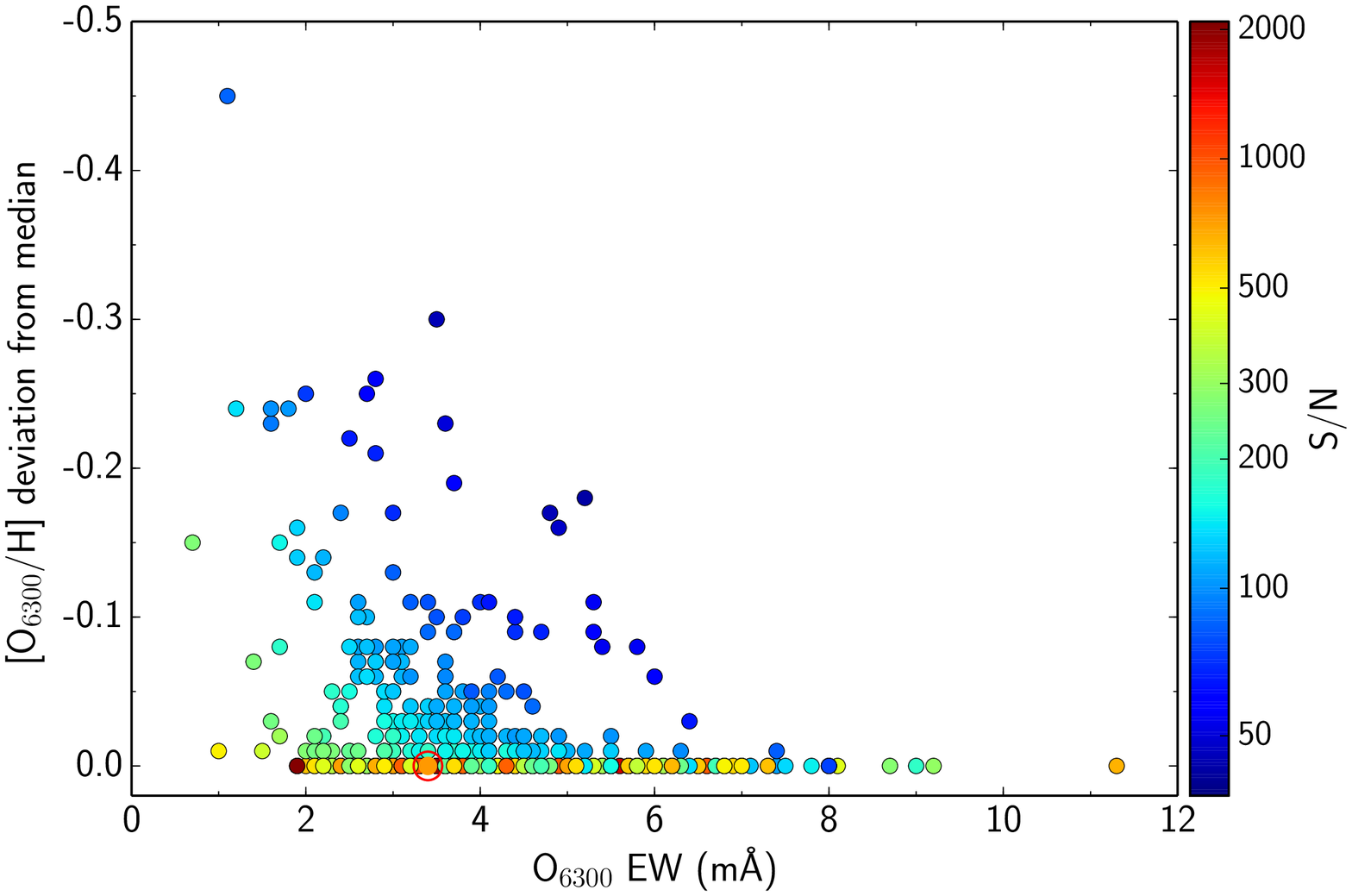}}
	\end{minipage} 
	\begin{minipage}[b]{0.49\linewidth}
		\resizebox{\hsize}{!}{\includegraphics{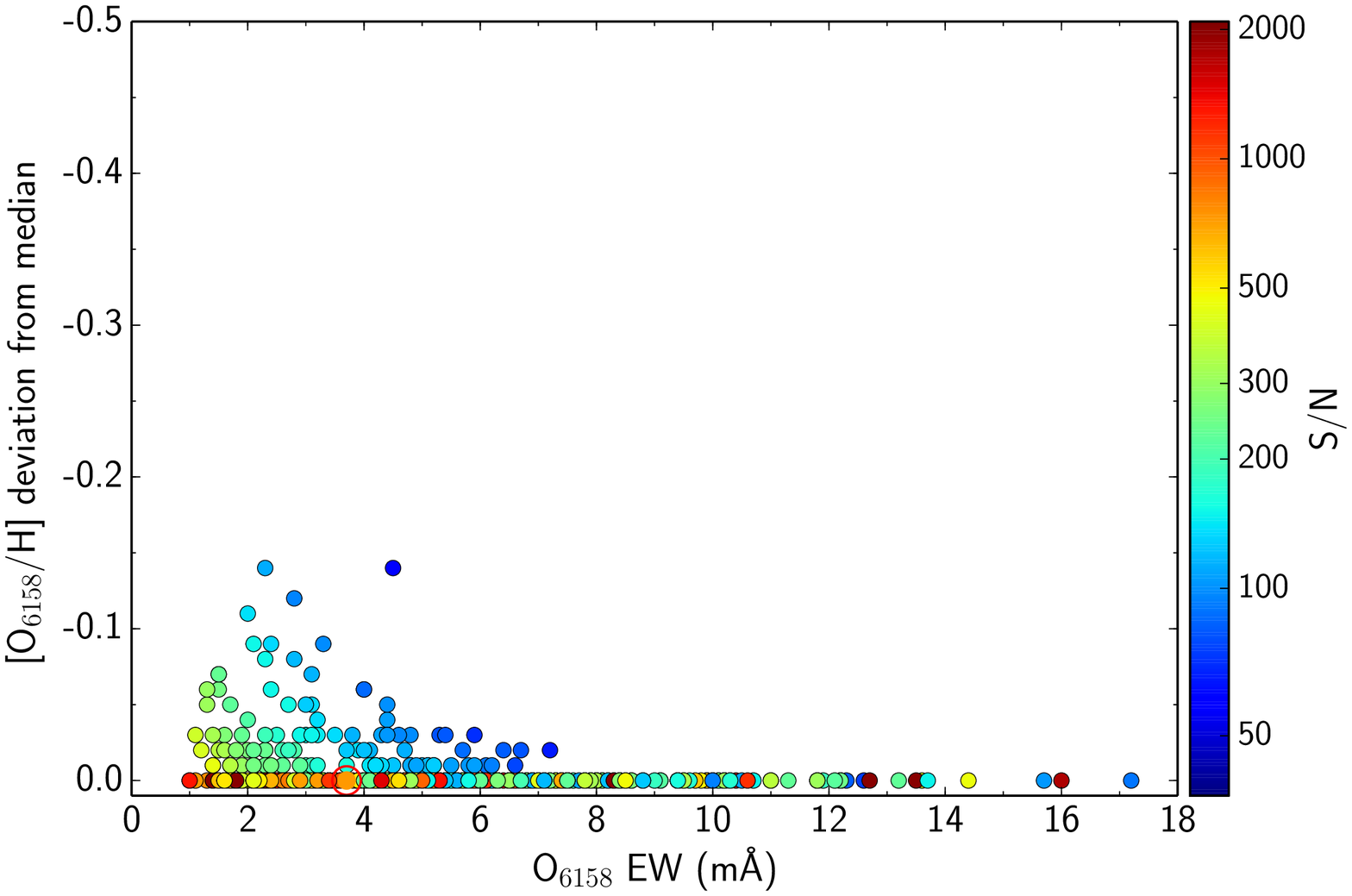}}
	\end{minipage} 
\caption{Deviation of the LTE abundances from the median value of the statistical probability distribution. Dependence on the nominal signal-to-noise is shown in colours.}
\label{DesvMedian}
\end{figure*}
% Two column figure 
\begin{figure*}[]
   \centering
	\begin{minipage}[b]{0.49\linewidth}
		\resizebox{\hsize}{!}{\includegraphics{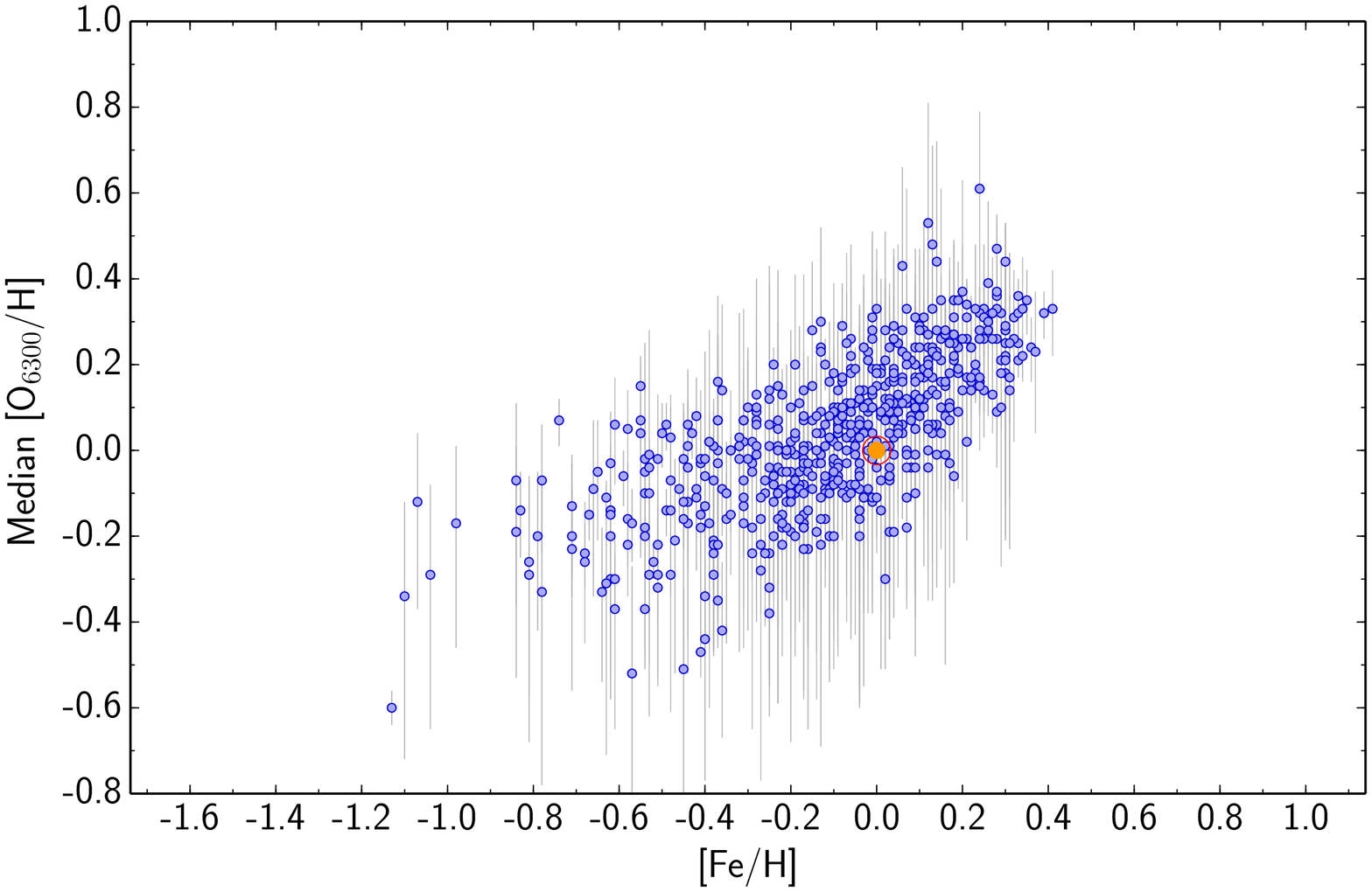}}
	\end{minipage} 
	\begin{minipage}[b]{0.49\linewidth}
		\resizebox{\hsize}{!}{\includegraphics{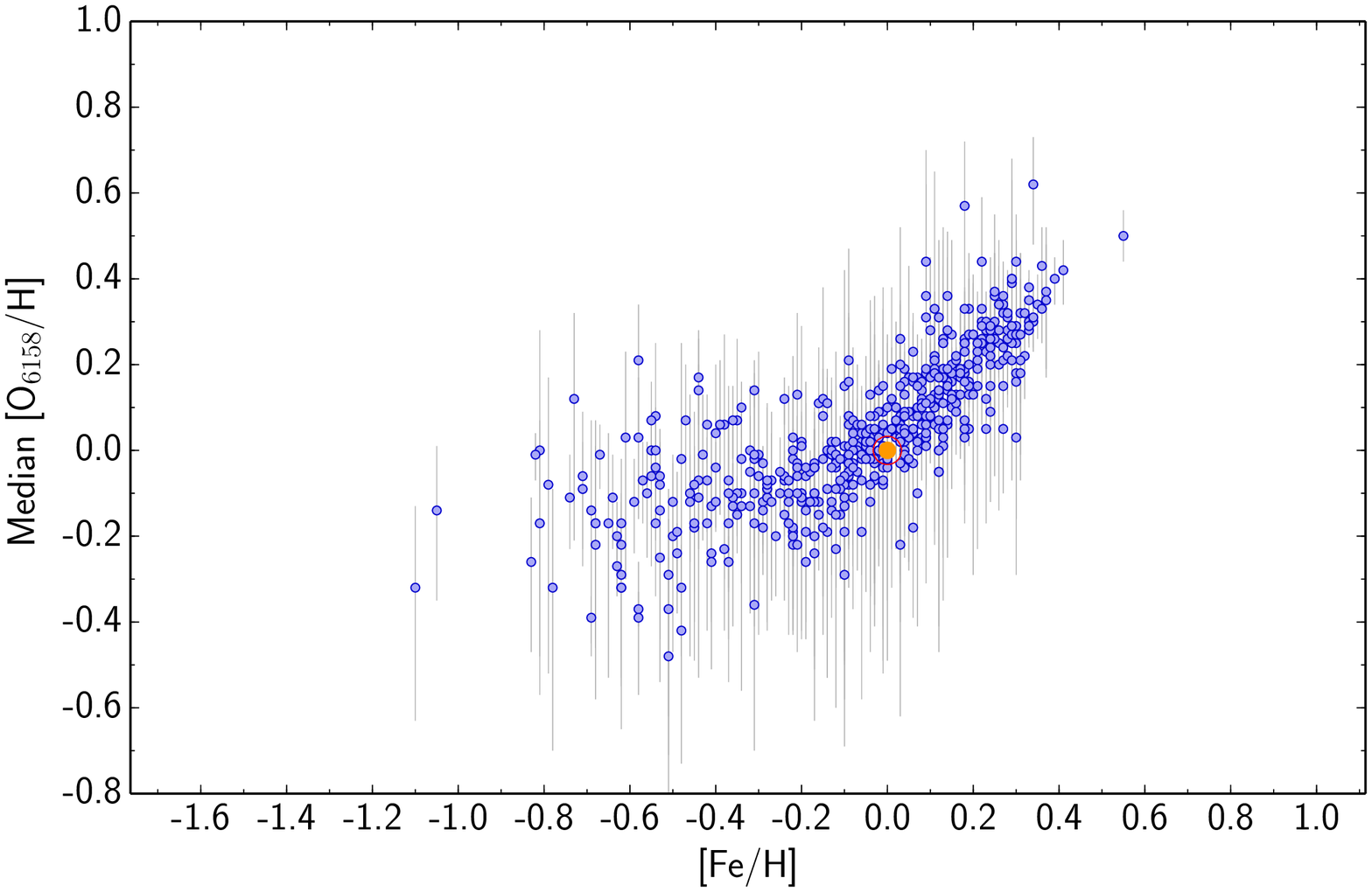}}
	\end{minipage} 
\caption{[O/H] vs. [Fe/H] for both oxygen abundance indicators, where [O/H] is taken as the median of the statistical distribution. The yellow dot corresponds to the solar abundances. [O/H] absolute error bars are over-plotted.}
\label{Errors}
\end{figure*}

\begin{table*}
\caption{Statistical oxygen abundances from the OI 6158 $\AA$ line}
\label{O6158Stats}
\centering
\begin{tabular}{lccccccc}
\hline\hline \\[-8pt]
 Star & EW$_{OI}$ (m$\AA$) & $\delta$EW$_{OI}$ (m$\AA$) & [O/H]$_{OI}$ & Median [O/H]$_{OI}$ & $\sigma_{OI}$ & +$\delta$[O/H]$_{OI}$ & -$\delta$[O/H]$_{OI}$ \\[2pt]
\hline \\[-8pt]
Sun&3.7&0.3& 0.00& 0.00&0.04&0.04&-0.05\\
HD70889&6.9&0.5& 0.10& 0.10&0.04&0.04&-0.04\\
HD21161&7.5&2.8& 0.18& 0.18&0.27&0.18&-0.26\\
HD37226&7.4&0.8&-0.01&-0.01&0.06&0.06&-0.06\\
...&...&...&...&...&...&...&...\\[2pt]
\hline
\end{tabular}
\tablefoot{The full table is available at the CDS.}
\end{table*}

\begin{table*}
\caption{Statistical oxygen abundances from the [OI] 6300 $\AA$ line}
\label{O6300Stats}
\centering
\begin{tabular}{lccccccc}
\hline\hline \\[-8pt]
Star & EW$_{[OI]}$ (m$\AA$) & $\delta$EW$_{[OI]}$ (m$\AA$) & [O/H]$_{[OI]}$ & Median [O/H]$_{[OI]}$ & $\sigma_{[OI]}$ & +$\delta$[O/H]$_{[OI]}$ & -$\delta$[O/H]$_{[OI]}$ \\[2pt]
\hline \\[-8pt]
Sun&3.4&0.3& 0.00& 0.00&0.04&0.04&-0.05\\
HD48611&1.9&0.8&-0.42&-0.42&0.26&0.16&-0.25\\
HD70889&3.3&0.4& 0.10& 0.10&0.05&0.05&-0.06\\
HD21161&5.9&2.9& 0.23& 0.24&0.29&0.17&-0.28\\
...&...&...&...&...&...&...&...\\[2pt]
\hline
\end{tabular}
\tablefoot{The full table is available at the CDS.}
\end{table*}

Because of the non-linearity of the transformation, this distribution is remarkably asymmetric. Several problems will arise from this fact, such as the asymmetry of the abundance errors or the biased result towards the high tail which is obtained from the LTE abundance analysis. 

Ideally, the normal distribution of the EW is not truncated, and the measured EW of the spectral line corresponds to the mean, median and mode, as expected for normal distributions. Under monotonic transformations the median is preserved and therefore the abundance from the LTE analysis will be the median of the distribution, acceptably quoted as a result, but no longer the mean or mode. However, when we address the measurement of small lines, the normal distribution is truncated at zero as a result of the negative values of EW. In this situation the measured EW is only the mode of the distribution (not the median), and thus its transformation will not yield the median of the abundance probability distribution: the abundance derived in the LTE analysis will not only be biased towards the left tail of the distribution, but will also be smaller than the median. In this case, it becomes especially useful to calculate the probability distribution in order to provide statistically consistent results. Abundances derived from the LTE analysis have been used in the present work; however, we provide in Tables~\ref{O6300Stats} and~\ref{O6158Stats} the statistical results. Figure~\ref{DesvMedian} shows the deviation of the LTE abundances from the median value of the statistical distribution. It is shown that for EWs larger than 8 m\AA~, regardless of the signal-to-noise, the result from the LTE analysis coincide with the median of the abundance statistical probability distribution. 

Abundance uncertainties are often calculated separately and added together in quadrature (positive and negative contributions independently). Nevertheless, this procedure is not justified mathematically \citep{Barlow03}. Throughout this paper we provided the errors in EW, which are symmetric. However, the uncertainties in abundance should be calculated using the probability distribution. The errors are calculated as the abundance values where the probabilities are 0.159 and 0.841 for the negative and positive errors, respectively. These abundance intervals would enclose the $\pm$1 $\sigma$ region around the median, in a normal distribution. We present in Fig.~\ref{Errors} the median abundance with the corresponding uncertainties. Tables~\ref{O6300Stats} and~\ref{O6158Stats} show abundance results from this statistical analysis for the OI6158$\AA$ and [OI]6300$\AA$ lines, respectively.

\end{document}